\documentclass[12pt,onecolumn,showpacs,amssymb,aps,nofootinbib,floatfix]{revtex4-1}
\usepackage{graphicx,epsfig,epstopdf}
\usepackage{xcolor}
\usepackage{amssymb}
\usepackage{amsmath}
\newcommand{\figref}[1]{Fig. \ref{#1} }
\newcommand{\secref}[1]{section \ref{#1} }
\newcommand{\appref}[1]{Appendix \ref{#1} }
\newcommand{\ave}[1]{\left\langle #1 \right\rangle}

\newcommand{\eqcomma}{\phantom{AA},\phantom{AA}}

\newcommand{\changed}[1]{\textcolor{black}{#1}}
\begin{document}
\title{Quarkonium spin alignment in a vortical medium}
\author{Paulo Henrique De Moura$^1$,Kayman J. Gonçalves$^1$, Giorgio Torrieri$^{1,2}$}
\affiliation{$\phantom{A}^1$ Universidade Estadual de Campinas - Instituto de Fisica "Gleb Wataghin"\\
Rua Sergio Buarque de Holanda, 777\\
 CEP 13083-859 - Campinas SP\\
$\phantom{A}^2$ 
Institute of Physics, Jan Kochanowski University, Ul. Uniwersytecka 7, 25-406 Kielce, Poland.}
\begin{abstract}
We use a potential model to investigate the phenomenology of quarkonium \changed{in} a thermal rotating medium, where vorticity and spin density are not necessarily in equilibrium.   We find that the quarkonium spin density matrix, as well as the binding energy and melting temperature, are sensitive to both the vorticity and the lack of equilibrium between vorticity and spin.    This means that quarkonium spin alignment is a sensitive probe for vorticity and spin within the hydrodynamic phase.
Information unequivocally pointing to spin-orbit non-equilibrium dynamics can be obtained from a combined study of quarkonium relative abundance and spin alignment, as well as experimentally obtainable off-diagonal density matrix elements.  
\end{abstract}

\maketitle
\section{Introduction}
Quarkonium as been used as a probe of thermodynamic properties of the Quark-Gluon Plasma (QGP) since the seminal paper from \cite{MATSUI_SATZ}.  The heavy mass scale means that it is a probe that can both be examined reliably by theoretical calculations and susceptible to non-trivial in-medium effects, serving as a ``thermometer'', sensitive to the interplay of thermal fluctuations and the QCD medium, and as a probe for non-equilibrium behavior \cite{review,classic1,classic2,classic3}.

The advent of the study of vorticity in heavy ion collisions \cite{bec} added a potentially new arena where quarkonium could be used.   In fact, we shall argue that quarkonium provides unique opportunities for the phenomenology of hydrodynamics with spin.      First of all, quarkonium can be formed early in the collision and can survive throghout the quark gluon plasma evolution.   Unlike polarized $\Lambda$s \cite{bec} and spin-aligned vector mesons \cite{alices,star,wang,xia,kayman,oliva}, they are potentially sensitive to the entire dynamics of the fluid and not just to the freeze-out.

Furthermore, the long-lived quarkonium state is spin 1, having a 3$\times$3 density matrix with 8 degrees of freedom, 6 of which are accessible to spin alignment measurements.   Thus, if, as seems to be theoretically highly likely \changed{\cite{uscausality,uscas2,usdiss,relax,jeon,gursoy,hongo,hongo2}}, spin and vorticity are not in equilibrium, this lack of equilibration will can be imprinted on the density matrix's measurable off-diagonal elements $\rho_{0,\pm 1},\rho_{\pm 1,\mp 1}$ \cite{kayman} (No equivalent elements exist for the ``qubit'' density matrix of a fermion).  In fact, these elements have recently been measured \cite{ALICE}.

Last but not least, bottomonium and to a certain extent charmonium states can be viewed as solutions to a Schrodinger equation with heavy Quark wavefunctions moving around a QCD potential (including a weakly coupled and a confined part) \cite{brambilla,boundquark}.  It is therefore possible to understand, both analytically \cite{tuchin} and via effective theory \cite{kim}, effect that vorticity will have to the properties of quarkonium.

In this work, we shall go in this direction, combining the insights developed in \cite{kayman} with a potential model for quarkonium solved using standard methods extended to rotating frames \cite{NU,ARF,Anan,Anan2} and finite temperature \cite{AHM,WONG}.  In \secref{prelim} we shall assess currently available experimental data, and point out what would be necessary to probe \changed{spin-orbit} non-equilibrium.  Then in \secref{potential} we calculate the quarkonium wavefunction properties, namely binding energy and a semi-classical estimate of the melting temperature, in a rotating frame.  Finally, in \secref{offdiag} we calculate quarkonium observables that could indicate a lack of equilibrium between spin and vorticity.
\section{The Quarkonium spin density matrix \label{prelim}}
%We will use natural notation in this paper $\hbar = 1$ and $c=1$. 

In \cite{kayman}, we have argued that vector spin alignment contains crucial information on the still-unknown spin hydrodynamic evolution in heavy ion collisions, provided that not just the $\rho_{00}$ coefficient but the ``off-diagonal'' coefficients are measured.
The former, $\rho_{00}$ was measured in \cite{alices,star} and is sensitive to $\theta$, the angle w.r.t. the spin alignment direction, which for heavy ions would be mainly the reaction plane, reflecting the vorticity structure in analogy to the global polarization measurement \cite{bec}.

The other coefficients, related to density matrix elements $\rho_{0\pm 1,\pm 1\pm 1,\pm 1\mp 1}$ (see equation (2)  of \cite{kayman}) would depend on a second ``reference'' angle $\phi$, whose most logical definition is in terms of the beam axis.

\begin{figure}[t]
		\begin{center}
			\includegraphics[scale=0.7]{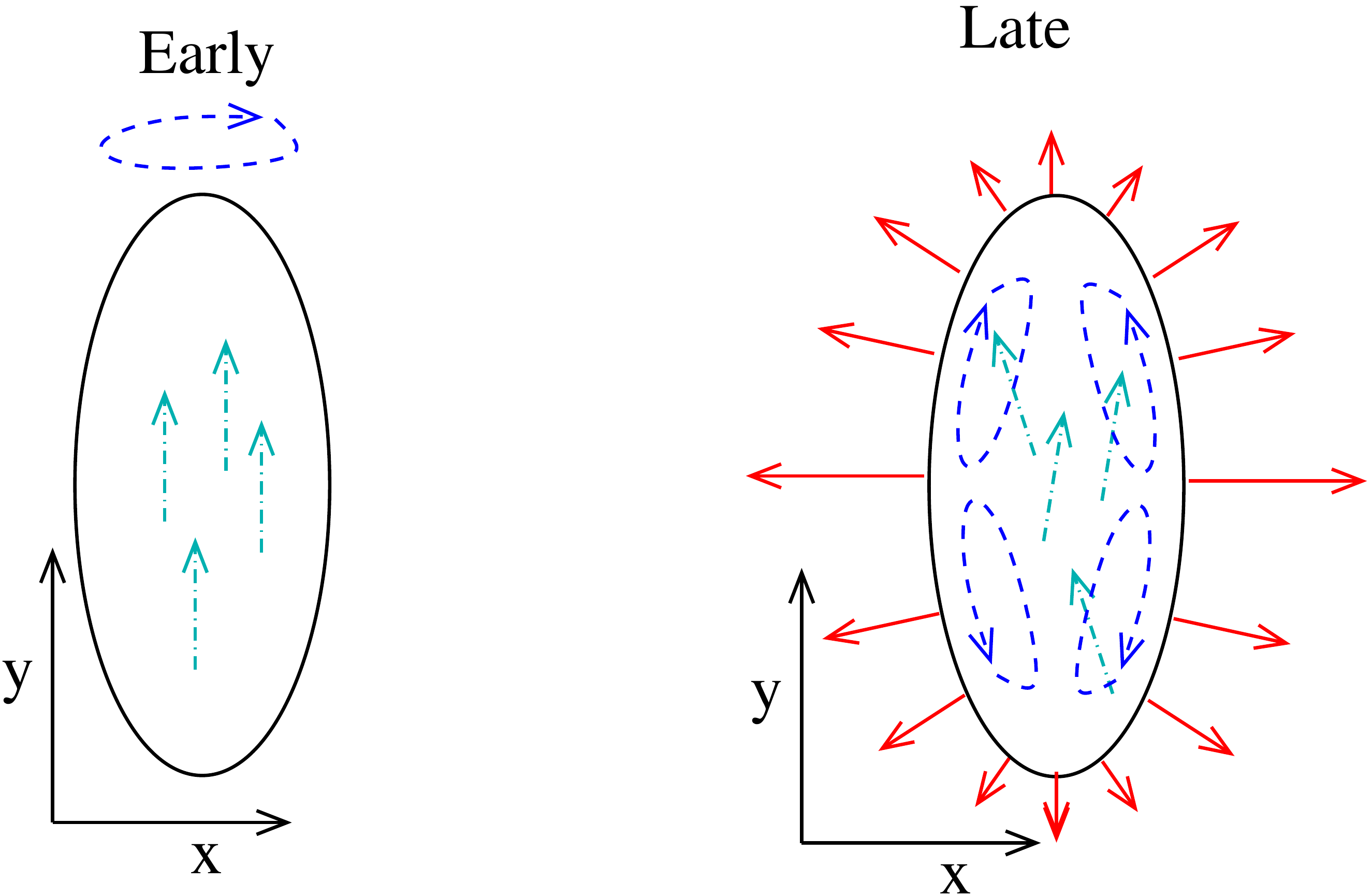}
			
\end{center}
                \caption{ A schematic representation of how the vorticity in heavy ion collisions looks like at early and late times \label{translong}
                Blue dashed arrows represent vorticity, cyan dot-dashed ones polarization, red solid ones flow.  The mis-alignment of spin in the right panel is due to the lack of equilibrium between vorticity and polarization \cite{uscausality,usdiss}.   }
	\end{figure}

In such a set-up, the non-equilibrium between spin and vorticity would manifest itself by the interplay of transverse vs. longitudinal polarization~\cite{bec}.  The heavy ion system actually lends hope that this interplay could occur (\figref{translong}):
Transverse polarization is thought to be present from the beginning, since it is the result of the angular momentum of the initial state of an off-central collision.
Longitudinal polarization forms from azimuthal gradients, on a time-scale comparable to the formation of momentum anisotropy causing elliptic flow.   If polarization ``relaxes to vorticity'' \cite{uscausality,usdiss}, one expects polarization to be aligned in the transverse direction, but comparatively mis-aligned in the longitudinal one \cite{usdiss}, which indeed seems to be out of phase with some dynamical calculations \cite{bec}.    \textcolor{black}{Hence, there will be an angle in the $z-y$ plane between vorticity and polarization, whose magnitude we do not yet know to estimate quantitatively but whose experimental confirmation will add a phenomenological dimension to the recent  theoretical development of spin hydrodynamics \cite{uscausality,uscas2,usdiss,relax,jeon,gursoy,hongo,hongo2}}.

The hope, therefore, is that off-diagonal (longitudinal-transverse) coefficients are potentially important for they could signal deviations from local equilibrium (included via a Cooper-Frye type formula \cite{zanna}) due to the presence of two distinct axial currents, representing spin and vorticity, evolving on different time-scales \cite{usdiss}. In \cite{kayman} we illustrated this with a coalescence type model.

Coalescence of only spin within a vortical background should not change the coherence of the density matrix, since it is a unitary process and the dynamics is symmetric around the vortical axis.  But assuming vorticity and pre-existing spin density are not in equilibrium and pointing in different directions, this is no longer true \cite{kayman};
Vorticity is ``classical" background, interacting with the quantum spin state, so if coalescence happens in a vortical background (i.e. if spin and vorticity are out of equilibrium) one expects impurity of the density matrix.
Mathematically, the loss of purity is manifest in Eq. (23) of \cite{kayman} $P_L(w)$ representing the (unknown) classical probability of a vortex $\omega$ giving an angular momentum $L$ to the meson wavefunction.   When this probability becomes uniform ($P_L(w) \rightarrow$ constant ) we recover a maximally impure state\footnote{Through not quite the Cooper-frye ansatz of \cite{zanna}. A $P_L=$ constant impure state can be regarded as a microcanonical density matrix assuming the diquark quarkonium state is exact.  The grand canonical matrix inherent in the Cooper-Frye formula of \cite{zanna} and it's vector extension would arise if all values of $L$ up to $\infty$ were allowed due to angular momentum fluctuations and a bath of degrees of freedom.}.  For vector bosons, this impurity is manifest in the off-diagonal matrix components (see the discussion between eqs 4 and 5 of \cite{kayman}).

While this data as yet does not exist for vector mesons, it does exist for quarkonia $J/\psi$ and $\Upsilon$ states \cite{ALICE}, since the ALICE Collaboration measurement of the quarkonium polarization included the off-diagonal values of the spin density matrix.   Therefore, we can do direct connection between polarization parameters $\lambda_{\theta}$,  $\lambda_{\phi}$ and $\lambda_{\theta\phi}$ and density matrix \cite{ALICE} and the coefficients used in \cite{kayman}  (table in eq. \ref{coefftab})
  \begin{equation}
  \begin{array}{ccc}
\mathrm{Variable} &    \mathrm{Element}& \mathrm{coefficient}\times \frac{3}{4\pi}\\
 \rho_{00} &    \rho_{00} & \cos^2\theta\\
 \frac{1-\rho_{00}}{2} &   \frac{\rho_{11}+\rho_{-1-1}}{2} & \sin^2 \theta\\
    r_{10} & Re[\rho_{-10}-\rho_{10}] & \sin(2\theta)\cos(\phi)\\
   \alpha_{10} &   Im[-\rho_{-10}+\rho_{10}] & \sin(2\theta)\sin(\phi)\\
   r_{1,-1} &     Re[\rho_{1,-1}] & \sin^2 \theta \cos (2\phi)\\
   \alpha_{1,-1} &     Im[\rho_{1,-1}] & \sin^2 \theta \sin(2\phi)
  \end{array}
  \label{coefftab}
\end{equation}
  \begin{equation}
\label{charmcoeff}
\rho_{00}=\frac{1+\lambda_{\theta}}{3+\lambda_{\theta}} \eqcomma r_{1,-1}=\frac{\lambda_{\phi}}{3+\lambda_{\theta}} \eqcomma  r_{10}=\frac{\lambda_{\theta\phi}}{3+\lambda_{\theta}}
\end{equation}
This is possible by making the comparison between $J/\psi$ angular distribution and the standard vector mesons angular distribution shown in Eq. 1 of \cite{kayman} (Eq. \ref{charmcoeff}).

Thus, we can do the  analyses presented in \cite{kayman} to relate $\lambda_{\theta,\phi,\theta\phi}$ to the wave function coherence via the parametrization in terms of \changed{Gell-Mann} matrices. Choosing the $n_{3-8}$ basis for this parametrization, we need to solve the following system of algebraic equations derived in \cite{kayman} in terms of the frame relating the lab to the spin direction (defined by angles $\theta_r,\phi_r$)
\begin{eqnarray}
\frac{1}{12} \left(3 \left(n_8-\sqrt{3}\;
n_3\right) \cos \left(2 \theta _r\right)-\sqrt{3}\;
n_3+n_8+4\right)=\rho_{00} \label{sys1}\\[2mm]
\frac{\left(n_8-\sqrt{3} \;n_3\right) \sin
	\left(\theta _r\right) \cos \left(\theta _r\right) \cos
	\left(\phi _r\right)}{\sqrt{2}}=r_{10} \label{sys2}\\[2mm]
-\frac{\left(\sqrt{3}\;n_3+3 n_8\right) \sin
	\left(\theta _r\right) \sin \left(\phi _r\right)}{3
  \sqrt{2}}=\alpha_{10} \label{sys3}
\end{eqnarray}
\begin{equation}
\label{phirdef}
  \phi_r=-\frac{1}{2}\tan^{-1}\left(\frac{\alpha_{1,-1}}{r_{1,-1}}\right)
\end{equation}
 
Now, we will do the follow change variable $\tilde{n} = n_{8} - \sqrt{3}\; n_{3} $, and knowing that variables are equal to zero $\alpha_{10}$ and $\alpha_{1,-1}$. So, we can write this system equation the following form:
\begin{eqnarray}
\frac{1}{12} \left(3 \tilde{n}\cos \left(2 \theta _r\right)+\tilde{n}+4\right)=\rho_{00} \label{sys1}\\[2mm]
\frac{\tilde{n} \sin\left(\theta _r\right) \cos \left(\theta _r\right) \cos\left(\phi _r\right)}{\sqrt{2}}=r_{10} \label{sys2}\\[2mm]
\phi_r = 0 \label{phirdef}\\
\end{eqnarray}
Therefore, we have the following  solution: 

\begin{eqnarray}
	\tilde{n}\left(\rho_{00},r_{10}\right) =-\frac{(1-3 \rho_ {00})^2+3 \sqrt{(1-3 \rho_{00})^4+4 (1-3 \rho_{00})^2 r_{10}^2}}{6 \rho_{00}-2}\label{ntilde}\\[3mm]
   \Theta_1\left(\rho_{00},r_{10}\right) = -\sqrt{\frac{2 (1-3 \rho_{00})^2-2 \sqrt{(1-3 \rho_{00})^4+4 (1-3 \rho_{00})^2 r_{10}^2}+6 r_{10}^2}{2 (1-3 \rho_{00})^2+9
   r_{10}^2}} \\[3mm] 
   \Theta_2\left(\rho_{00},r_{10}\right) = \sqrt{\frac{(1-3 \rho_{00})^2-\sqrt{(1-3 \rho_{00})^4+4 (1-3 \rho_{00})^2 r_{10}^2}+3 r_{10}^2}{2 (1-3 \rho_{00})^2+9 r_{10}^2}}\\[3mm]
   \Theta_3\left(\rho_{00},r_{10}\right) = \frac{ \Theta_2\left(\rho_{00},r_{10}\right)\left(2 (1-3 \rho_{00})^2+2 \sqrt{(1-3\rho_{00})^4+4 (1-3 \rho_{00})^2 r_{10}^2}\right)}{2 (3 \rho_{00}-1) r_{10}}\\[3mm]
   \theta_r = \tan ^{-1}\left(\Theta_1\left(\rho_{00},r_{10}\right), \Theta_3\left(\rho_{00},r_{10}\right)\right)
\end{eqnarray}
So, using polarization parameters that were obtained from ALICE collaboration \cite{ALICE} at different transverse momentum ($p_{T}$) ranges, we can determine whether the density matrix represents a coherence state or not. To make it we will use the equation \ref{ntilde} and reach the following figure \ref{states}.
\begin{figure}[t]
		\begin{center}
			\includegraphics[scale=0.7]{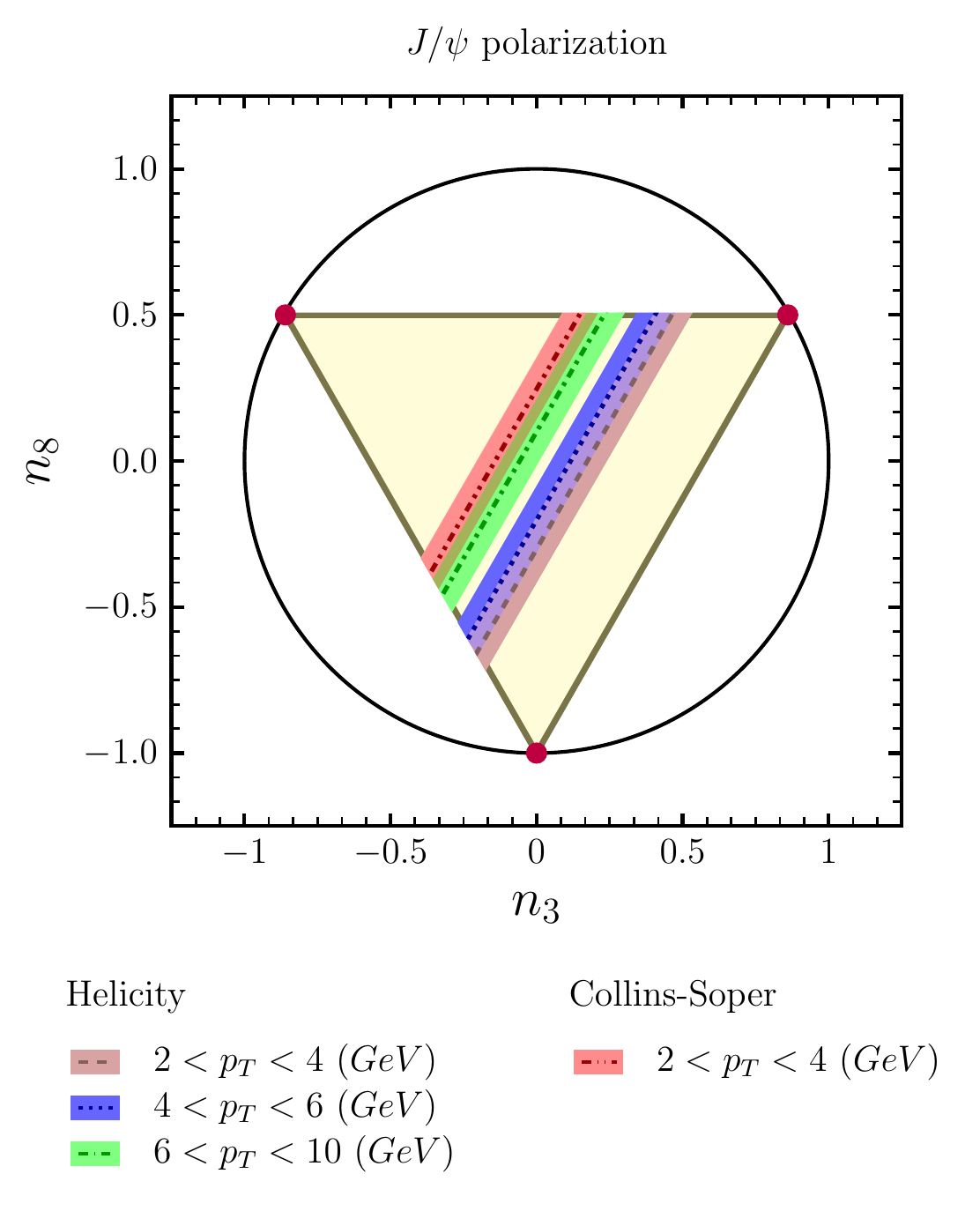}
			\includegraphics[scale=0.7]{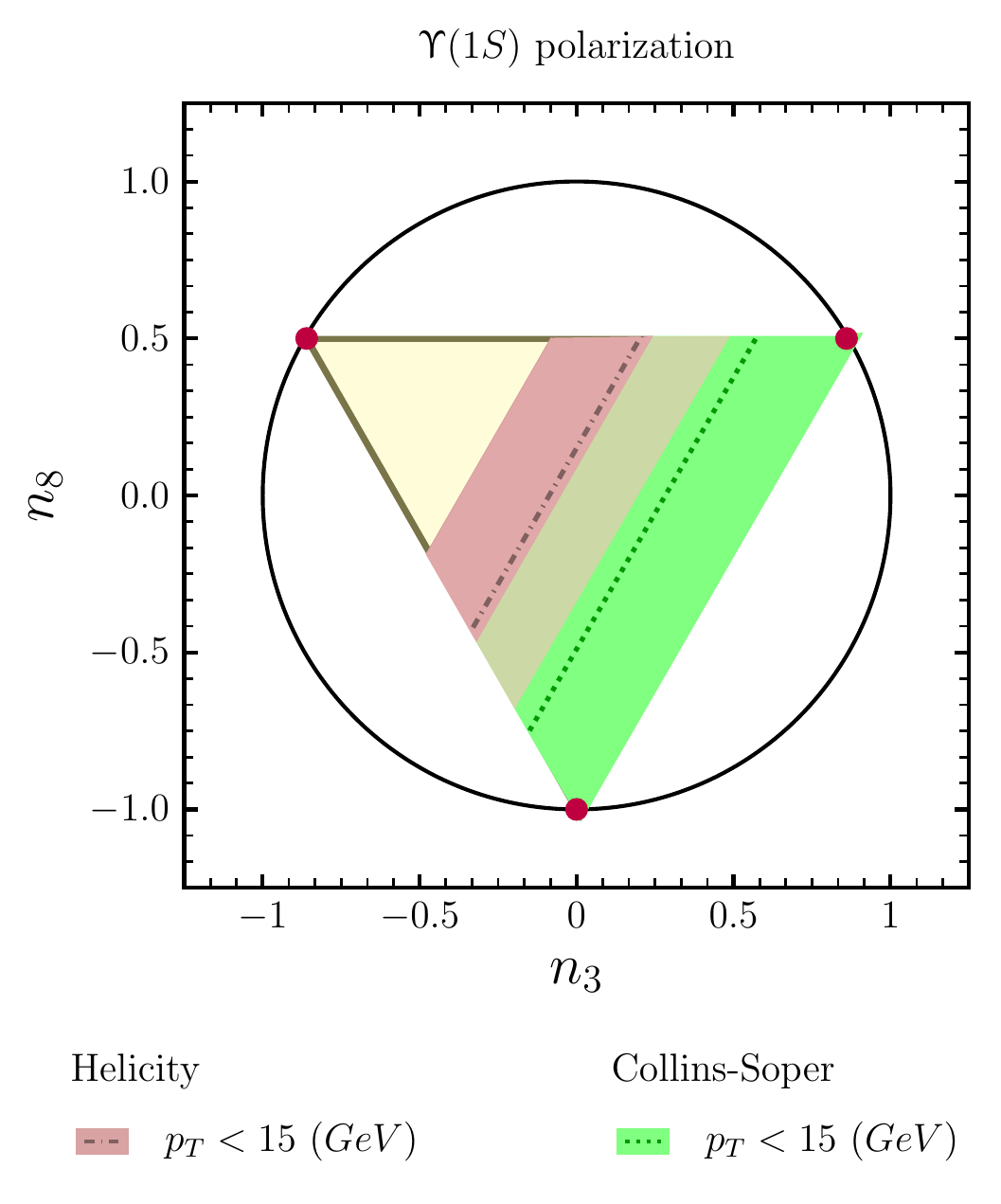}
\end{center}
                \caption{The experimental results of \changed{$J/\psi$} and $\Upsilon$ polarization measurements analyzed in terms of the \changed{Gell-Mann} matrix representation of \cite{kayman}.
                  These results were made using the helicity and Collins-Soper frame to $J/\psi$ polarization \cite{ALICE}. The other Collins-Soper values was not take in this plot because they were above the Helicity $6 < p_T < 10 ~ (GeV/C^2)$ uncertainty band as were shown in the text.   \label{states}}
	\end{figure}

Now, the coefficients in the frame Collins-Soper frame, given in the ranges $4 < p_T < 6 ~ (GeV/C^2)$ and $6 < p_T < 10 ~ (GeV/C^2)$ respectively result in $\tilde{n}=(0.09\pm 0.10, 0.02\pm 0. 07)$. \changed{Making} the comparison with the Helicity frame value $\tilde{n}=0.09\pm 0.11$ to $6 < p_T < 10 ~ (GeV/C^2)$. Therefore,  we can see that they are the same within error bar.

Looking at the figure \ref{states}, we can conclude that the density matrix from $J/\psi$ particle does not represent the pure state since none of the values for $n_{3,8}$ obtained from the data intersects the black pink points, i.e. these points represent the pure state in other words when the density matrix satisfy $\rho^2 = \rho$. This might indicate that statistical freeze-out advocated in \changed{\cite{zanna,classic1}} is a good estimate of particle production in heavy-ion collisions because the density matrix does not represent a coherent state as argued in \cite{kayman}.
In the bottomonium $\Upsilon (1S)$ case, we can see in the right panel of \figref{states} that because of large uncertainty do not know whether for these particles the density matrix represents a coherence state or not. 

\textcolor{black}{To fully assess the significance of the estimate above one must recall  Fig. 4 of \cite{kayman} and the definition of $P_L(w)$ given in the introduction}.
We have no idea what $P_L(w)$ is beyond the fact that it overall conserves momentum, but it acts as a projector.   One recovers a pure state when $\rho_{ij}=\delta_{Ll}$, (there is a certainty of vorticity giving a certain momentum) and a maximally mixed state when the momentum given by vorticity is independent of $L$. So the measurement in Fig. 4 \cite{kayman} is directly connected to how out of equilibrium vorticity and spin are, and how much vorticity vs pre-existing spin influences the final spin of the vector meson.  Linear combination of the different $L$-values in Fig. 4 of \cite{kayman} are possible, illustrating a probability of different spin configurations.

Note that these coefficients are given in terms of an angle $\theta$, which in \cite{kayman} is related to $\theta_r$, the angle between the hadronization frame and the lab frame.   This angle of course depends on the detailed hydrodynamical and spin-hydrodynamical evolution of the system, but it is obviously highly dependent on the reaction plane angle $\Phi$.   \textcolor{black}{Note that since local polarization necessarily averages to zero in symmetric collisions by parity, so must the off-diagonal matrix elements (unless large event by event fluctuations in transparency occur, something never seen in experimental data).
  Because of this, a zero finding integrated over azimuthal anisotropy can not count as conclusive evidence. }

\begin{figure}[t]
		\begin{center} 
			\includegraphics[scale=0.45]{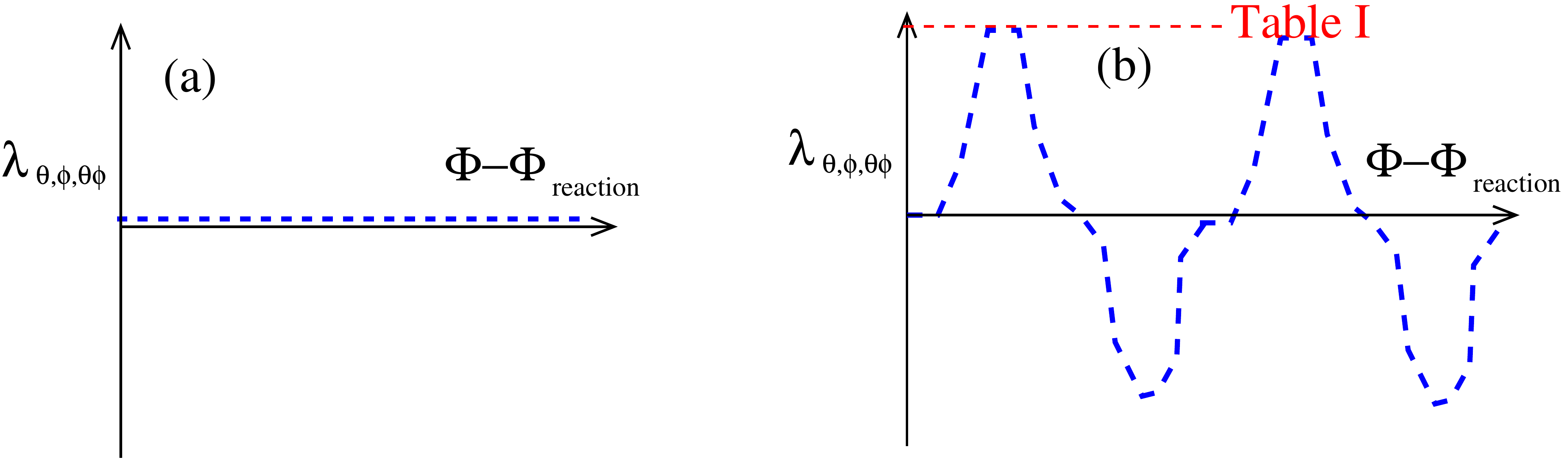}
			\label{figpl}
                        \caption{\label{pl} A schematic illustration of how we expect the coefficients \changed{$\lambda_{\theta,\phi,\theta \phi}$} to evolve with reaction plane angle $\Phi-\Phi_{reaction}$ in the two scenarios.   In \cite{ALICE}, these coefficients are compatible with zero when integrated over this angle.   A flat dependence (a) indicates a maximally incoherent Cooper-Frye/thermal type production.  A harmonic dependence (b) might indicate a more non-trivial coalescence scenario involving vorticity.  Note that both cases are consistent with zero when integrated over $\Phi$, as was done in \cite{ALICE}.
                          NB: The exact azimuthal shape will need more realistic modeling, but the amplitude can be qualitatively estimated in Table \ref{tab_quarkonium1} using \cite{blast1,blast2}}
        \end{center}
\end{figure}

Considering the Harmonic behavior of the coefficients in Fig. 4 of \cite{kayman} w.r.t. $\theta_r$ (most coefficients average to zero for all angles), therefore, it would be crucial to measure $\lambda_{\theta,\phi,\theta\phi}$ not as a function of $p_T$ as in \cite{ALICE} but as function of azimuthal reaction plane angle (\textcolor{black}{This can be measured directly at lower energies \cite{event1}, and via cumulants at the LHC \cite{event2}}).
A modulated behavior would be a clear signature for a non-trivial $P_L(w)$ which can then be harmonically decomposed into $L=0,1,2$ components of Fig 4 of \cite{kayman} to obtain information of the impact of spin vs vorticity in $J/\psi$ vs $\Upsilon$ hadronization.   If the dependence w.r.t. $\theta_r$ will be compatible with zero as it was for $p_T$ in each $\Phi$ bin, this is good evidence for a statistical Cooper-Frye freeze-out as in \cite{zanna,classic1}.   Schematically, these two alternatives are illustrated in \figref{pl}.

The quantitative details of \figref{pl} would require a hydrodynamic simulation with a hydrodynamic model where spin and vorticity are not in equilibrium, \textcolor{black}{given the interplay between spin, vorticity,radial flow and its anisotropies in the quarkonia $p_T$ and rapidity distributions.   We have used an analytically solvable model subject to ongoing work \cite{blast1,blast2} to provide an order of magnitude estimate of the maximum of the effect compatible with Fig. \ref{states}.  For reasons expanded on in the appendix this should be regarded as just that, an order of magnitude estimate, although it is gratifying that the possible off-diagonal terms are comparable in magnitude to $\lambda_\theta$}.

For this reason, in the rest of the paper we proceed to examine the microscopic dynamics of the charmonium state using a potential model, with a view of developing {\em quantitative} signatures of non-equilibrium between charmonium spin and vorticity.

%\begin{table}[!b]
%\changed{
% \begin{tabular}{|p{2cm}|p{2cm}|p{3cm}|p{2cm}|p{2cm}|p{3cm}|}
% \begin{tabular}{p{2cm}p{2cm}p{3cm}p{2cm}p{2cm}p{3cm}}
%\begin{tabular}{|c|c|c|c|}
%\cline{2-4}
%	\multicolumn{1}{c|}{}     & $\lambda_{\theta}$ & $\lambda_{\phi}$ & $\lambda_{\theta\phi}$      \\\hline
%	$J/\psi$  &  0.42      &  0.45         &  0.24							\\\hline
%	$\Upsilon$  &  0.78      &  0.78        &  0.03           					\\\hline
%\end{tabular}
	%\caption{\label{tab_quarkonium1} These are estimative of density matrix coefficients using Blast Wave Model \cite{blast1,blast2}} more deltails in Appendix.}
%}
%\end{table}
\begin{table}[!b]
\changed{
% \begin{tabular}{|p{2cm}|p{2cm}|p{3cm}|p{2cm}|p{2cm}|p{3cm}|}
% \begin{tabular}{p{2cm}p{2cm}p{3cm}p{2cm}p{2cm}p{3cm}}
\begin{tabular}{|c|c|c|c|}
\cline{2-4}
	\multicolumn{1}{c|}{}     & $\lambda_{\theta}$ & $\lambda_{\phi}$ & $\lambda_{\theta\phi}$  \\\hline
	$J/\psi$  &  0.021      &  0.023        &  0.012         		  \\\hline
	$\Upsilon$  &   0.054      &  0.054        & 0.002           		 \\\hline
\end{tabular}
	\caption{\label{tab_quarkonium1} \textcolor{black}{These are estimates of density matrix coefficients using Blast Wave Model \cite{blast1,blast2} more details in Appendix \appref{estim}.}}
}
\end{table}

\section{The quarkonium state  in rotating reference frames\label{potential}}
 \subsection{The Schr\"{o}dinger equation}
\changed{We are interested in the angular momentum due to vortices couples with the quark spin. So, we need to take an extra term, $\boldsymbol{\omega}\cdot\mathbf{S}$~\cite{Anan},}
\changed{
\begin{equation}
	\mathcal{H} = \frac{(\boldsymbol{p}-m\boldsymbol{\omega}\times\mathbf{r})^2}{2m} - \frac{m}{2}(\boldsymbol{\omega}\times\mathbf{r})^2 -\boldsymbol{\omega}\cdot\mathbf{S}+ V(r)
	\label{RotH}
\end{equation}}
Now, we can write the equation \ref{RotH} for a two-body case. Thus we have the following expression:
\changed{
\begin{equation}
	\mathcal{H} = \sum_{i=1,2}\left(\frac{(\boldsymbol{p}_i-m_i\boldsymbol{\omega}_i\times\mathbf{r}_i)^2}{2m_i} - \frac{m_{i}}{2}(\boldsymbol{\omega_{i}}\times\mathbf{r}_{i})^2 - \boldsymbol{\omega_{i}}\cdot\mathbf{S_i}\right) + V(\lvert \mathbf{r_1} - \mathbf{r_2} \rvert)
	\label{RotH2}
\end{equation}	 
}
Using the relations:

\begin{equation}
\left\{
\begin{array}{lcl}
    \mathbf{P} = \mathbf{p_1} + \mathbf{p_2},  & &
    \mathbf{p} = \mu \left(\dfrac{\mathbf{p_1}}{m_1} - \dfrac{\mathbf{p_2}}{m_2}\right) \\[3mm]
    \mu = \dfrac{m_1\;m_2}{m_1+m_2},  &\quad & \\[3mm]
    \mathbf{r_1} = \mathbf{R} +\dfrac{m_2}{m_1 + m_2}\mathbf{r}, & &
    \mathbf{r_2} = \mathbf{R} -\dfrac{m_1}{m_1 + m_2}\mathbf{r}
\end{array}
\right.
\end{equation}

As the two quarks that form the mesons are in the same vortical background, \changed{we can suppose that $\boldsymbol{\omega}_1 = \boldsymbol{\omega}_2 = \boldsymbol{\omega}$ and $\mathbf{S} = \mathbf{S_1} + \mathbf{S_2}$, and then write:}
\changed{
\begin{equation}
	\mathcal{H} = \frac{P^2}{2M} + \frac{p^2}{2\mu} - \mathbf{P}\cdot(\boldsymbol{\omega}\times\mathbf{R}) - \mathbf{p}\cdot(\boldsymbol{\omega}\times\mathbf{r})-\boldsymbol{\omega}\cdot(\mathbf{S}_{1}+\mathbf{S}_{2})+ V(\lvert \mathbf{r} \rvert) 
	\label{h2}	
\end{equation}
}
\changed{Since $\mathbf{S} = \mathbf{S}_{1}+\mathbf{S}_{2}$ and we are interested just in the reduced coordinate,} 
\changed{
\begin{equation}
	\mathcal{H} = \frac{p^2}{2\mu} - \mathbf{p}\cdot(\boldsymbol{\omega}\times\mathbf{r})-\boldsymbol{\omega}\cdot\mathbf{S}+ V(\lvert \mathbf{r} \rvert) 
\end{equation}
We can rewrite this equation as
\begin{equation}
	i\frac{\partial}{\partial t}= -\frac{1}{2\mu}\boldsymbol{\nabla}^2 - \omega L_z -\omega S_z+ V(\lvert \mathbf{r} \rvert) 
	\label{NonIn}	
\end{equation}}
\changed{
where we used $\mathbf{p} = - i\hbar \boldsymbol{\nabla}$, $\mathcal{H} = i\frac{\partial}{\partial t}$, $\boldsymbol{\omega} = \omega \mathbf{\hat{z}}$ and $L_z = -i  \frac{\partial}{\partial \phi}$.}

\changed{In rotating frames, the contribution in Hamiltonian is only the product between orbital angular momentum due to the vortices with the spin of meson. Thus, in this non-inertial frame, the contribution is just to $L$ different from zero. Then, we will write just the radial part of the equation \ref{NonIn}.}
\begin{equation}
	\frac{1}{r^2}\frac{d}{dr}\left(r^2\frac{dR(r)}{dr}\right) + \left[2\mu(E-V(r))-\frac{l(l+1)}{r^2}-2\mu\omega  L_z - 2\mu\omega S_z\right]R(r)=0	
	\label{shrr1}
\end{equation}
To go forward, we shall assume the rotation to be classical and related to the hydrodynamic vorticity.   Then we can define $\omega$ in terms of the a conserved  circulation

\begin{equation}
  \label{cir}
    C = \oint \mathbf{v}\cdot d\mathbf{l}
\end{equation}
and also assume the Cornell potential \cite{brambilla}

\begin{equation}
	V(r) = b\;r-\frac{\alpha_{eff}}{r} 
\end{equation}
so
\begin{equation}
	\frac{1}{2\mu r^2}\frac{d}{dr}\left(r^2\frac{dR(r)}{dr}\right) + \left[(E-V(r))-\frac{l(l+1)}{2\mu r^2} - \frac{\;m_j\; C}{\pi r^2}\right]R(r)=0
	\label{shrr2}
\end{equation}

Making $\chi (r) = r R(r)$ and change the variable $x=1/r$, then the equation \ref{shrr2}, we get:
\begin{equation}
        \frac{d^2\chi(x)}{dx^2}+\frac{2x}{x^2}\frac{d\chi(x)}{dx}+\frac{2\mu}{x^4}\left[E-\frac{b}{x}+\alpha_{eff}x-\frac{l(l+1)}{2\mu}x^2-\frac{m_j\;C}{2\pi}x^2\right]\chi (x) = 0
        \label{shrr3}
\end{equation}

Now we will expand the variable $y$ ($y = x-\delta$) around zero with $\delta = 1/r_0$ where $r_0$ is the mean meson radius. Therefore we have:
\begin{equation}
	\frac{b}{x}\approx b\left(\frac{3}{\delta}-\frac{3x}{\delta^2}+\frac{x^2}{\delta^3}\right)
\end{equation}
Thus we can write the equation \ref{shrr3} in following form:

\begin{equation}
        \frac{d^2\chi(x)}{dx^2}+\frac{2x}{x^2}\frac{d\chi(x)}{dx}+\frac{2\mu}{x^4}\left[E-\frac{3b}{\delta}+\left(\alpha_{eff}+\frac{3b}{\delta^2}\right)x-\left(\frac{l(l+1)}{2\mu}+\frac{m_j\;C}{2\pi}+\frac{b}{\delta^3}\right)x^2\right]\chi (x) = 0
\end{equation}

The coefficients $H_i$ with $i=0,1,2$, are given by:

\begin{equation}
        H_0 =  - 2\mu\left(E-\frac{3b}{\delta}\right)
\end{equation}

\begin{equation}
        H_1 =  2\mu\left(\alpha_{eff}+\frac{3b}{\delta^2}\right)
\end{equation}

\begin{equation}
        H_2 = -2\mu\left(\frac{l(l+1)}{2\mu}+\frac{m_j\;C}{2\pi}+\frac{b}{\delta^3}\right)
\end{equation}

Thus, we can write:
\begin{equation}
	\frac{d^2\chi(x)}{dx^2}+\frac{2x}{x^2}\frac{d\chi(x)}{dx}+ \frac{2\mu}{x^4}[-H_0+H_1\;x+H_2\;x^2]R(x)=0
\label{shro}
\end{equation}

Now, we can compare the differential equation \ref{shro} with \ref{hp}, we get:

\begin{equation}
		\sigma(x) = x^2 \eqcomma \tilde{\tau}(x) = 2x \eqcomma \tilde{\sigma}(x) = -H_0+H_1\;x+H_2\;x^2
\end{equation}

The function $\pi(s)$ from equation \ref{pi} is given by:

\begin{equation}
	\pi(x) = \pm \sqrt{(k-H_2)x^2-H_1 x+H_0}
\end{equation}
Choosing the negative solution the polynomial inside of square must have discriminant equal to zero, so we get:
\begin{equation}
	k = \frac{1}{4H_0}\left(H^2_1+4H_0H_2\right)
\end{equation}
Then,
\begin{equation}
	\pi(x) = -\frac{1}{2\sqrt{H_0}}\left(H_1 x - 2 H_0\right) 
	\label{tauaux}
\end{equation}
and $\tau(x)$ is given by equation \ref{tau}, so
\begin{equation}
	\tau(x) = 2x - \frac{1}{\sqrt{H_0}}\left(H_1 x -2 H_0\right) 
	\label{tau2}
\end{equation}
Then equaling the equations \ref{lambda1} and \ref{lambda2} we have:
\begin{equation}
	\frac{H^2_1}{4H_0} + H_2 - \frac{H_1}{2\sqrt{H_0}} = \frac{H_1}{\sqrt{H_0}} n - n(n+1)
	\label{h1eq}
\end{equation}
Solving the equation \ref{h1eq} to $H_0$, we get:

\begin{equation}
	\sqrt{H_0} = \frac{H_1}{(1+2n)\pm \sqrt{1-4H_2}}
\end{equation}
In this way, we can obtain the energy levels and wavefunctions as a function of the rotation parameters via the Eigenvalue equation
\begin{equation}
	\sqrt{-2\mu\left(E-\frac{3b}{\delta}\right)} = \frac{2\mu\left(\alpha_{eff}+\frac{3b}{\delta^2}\right)}{(1+2n)+\sqrt{1+4l(l+1)+\frac{4\mu\;m_j\;C}{\pi}+\frac{8\mu b}{\delta^3}}}
\end{equation}
\subsection{Mass and vorticity\label{masssec}}
The previous equation gives the binding energy in the rotating frame
\begin{equation}
	E_{n,l,m} = \frac{3b}{\delta} - \frac{2\mu\left(\alpha_{eff}+\frac{3b}{\delta^2}\right)^2}{\left[(1+2n)+\sqrt{1+4l(l+1)+\frac{4\mu\;m_j\;C}{\pi}+\frac{8\mu b}{\delta^3}}\right]^2}
	\label{energy}
\end{equation}
With he wave function determination is explicitly given in \appref{solution} using the method outlined in \appref{method}.
We can obtain the quarkonium mass from the equation \ref{energy}:
\begin{equation}
    M = 2m_q + E_{n,l,m}
\end{equation}
\begin{equation}
    M = 2m_q + \frac{3b}{\delta} - \frac{2\mu\left(\alpha_{eff}+\frac{3b}{\delta^2}\right)^2}{\left[(1+2n)+\sqrt{1+4l(l+1)+\frac{4\mu\;m_j\;C}{\pi}+\frac{8\mu b}{\delta^3}}\right]^2}
\end{equation}

\begin{figure}[h]
    \includegraphics[scale=0.6]{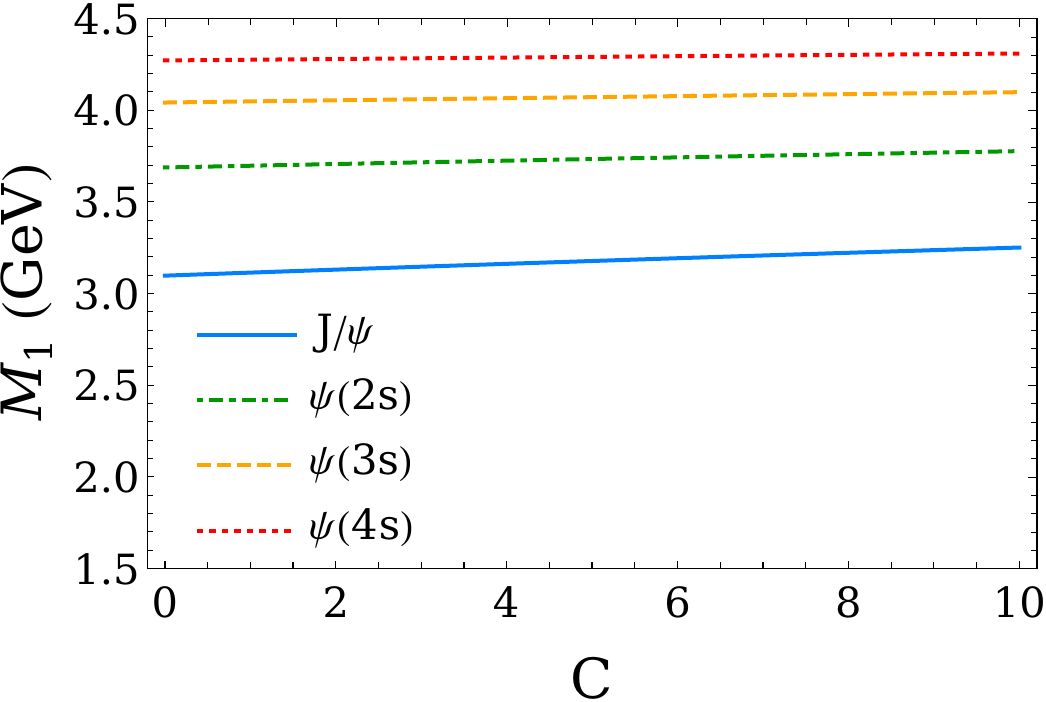}
    \includegraphics[scale=0.6]{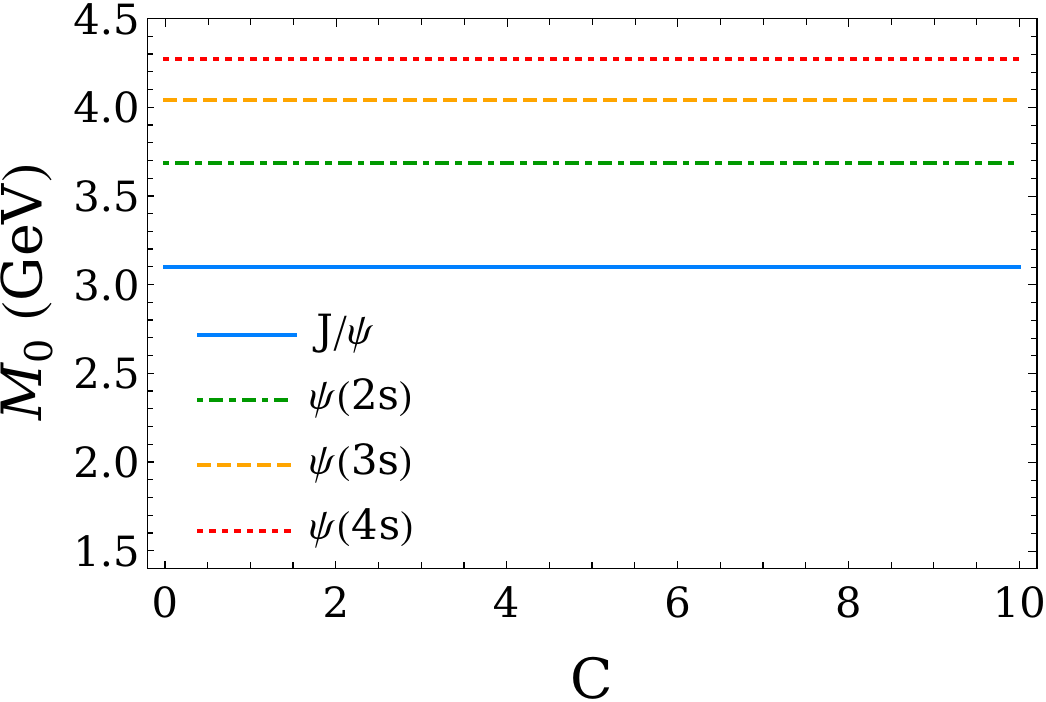}
    \includegraphics[scale=0.6]{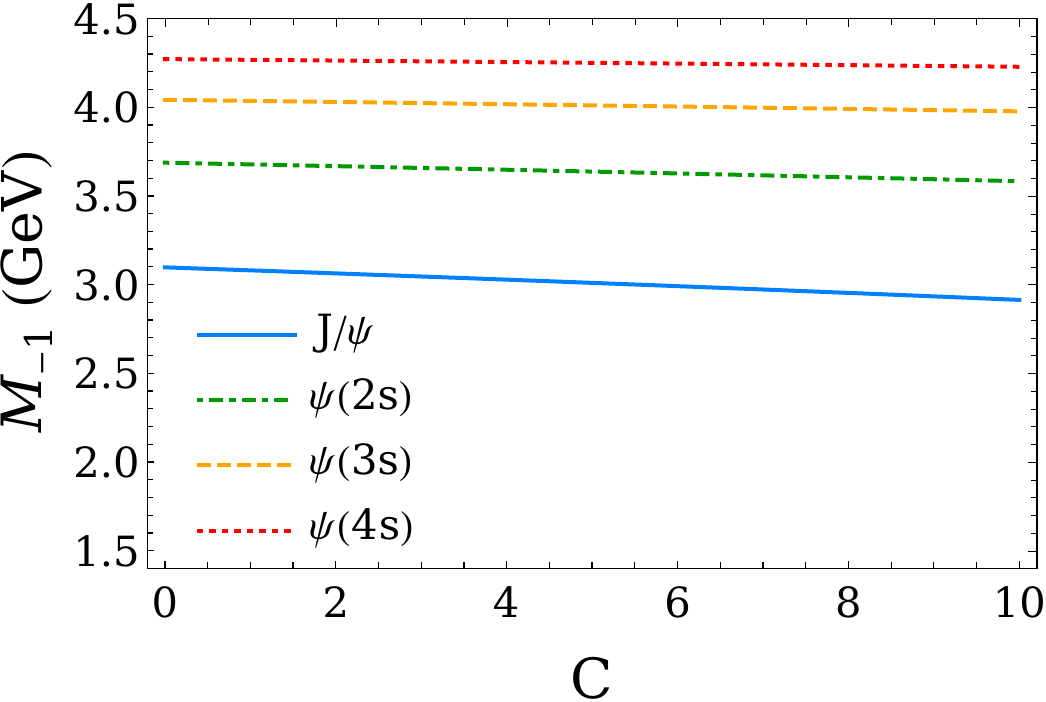}
    \includegraphics[scale=0.6]{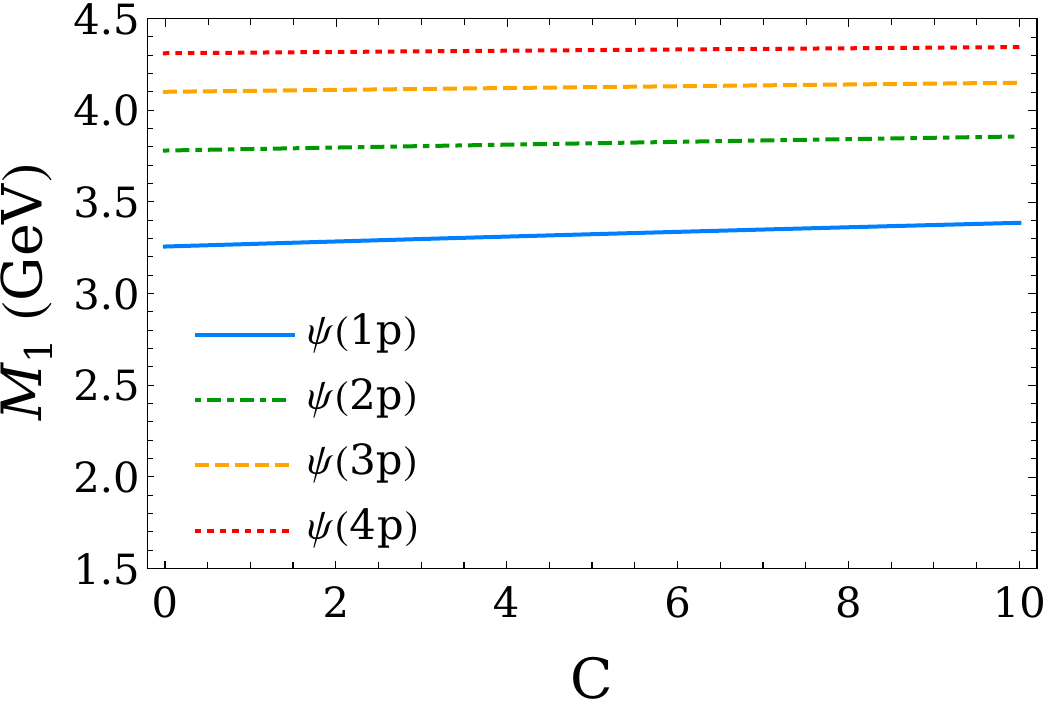}
    \includegraphics[scale=0.6]{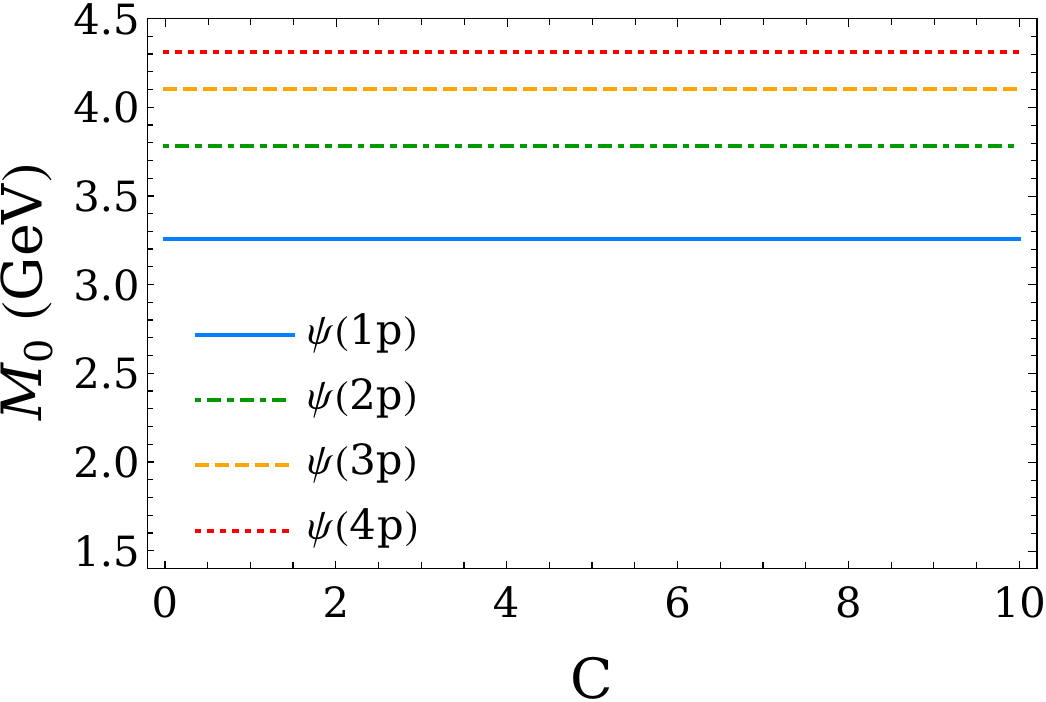}
    \includegraphics[scale=0.6]{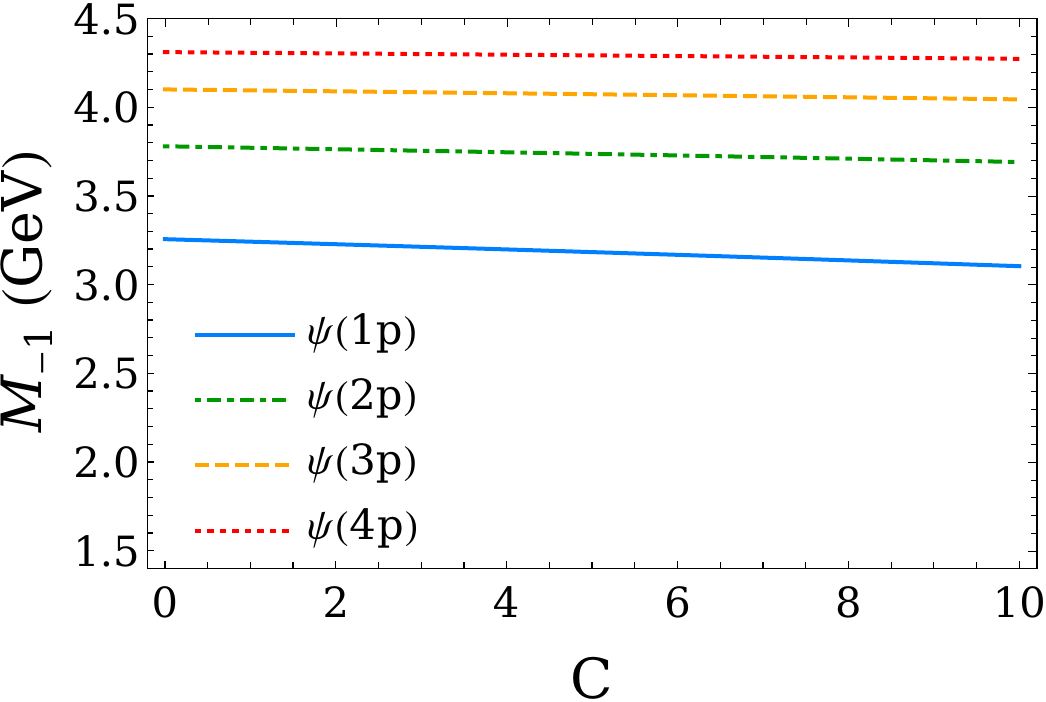}
    \includegraphics[scale=0.6]{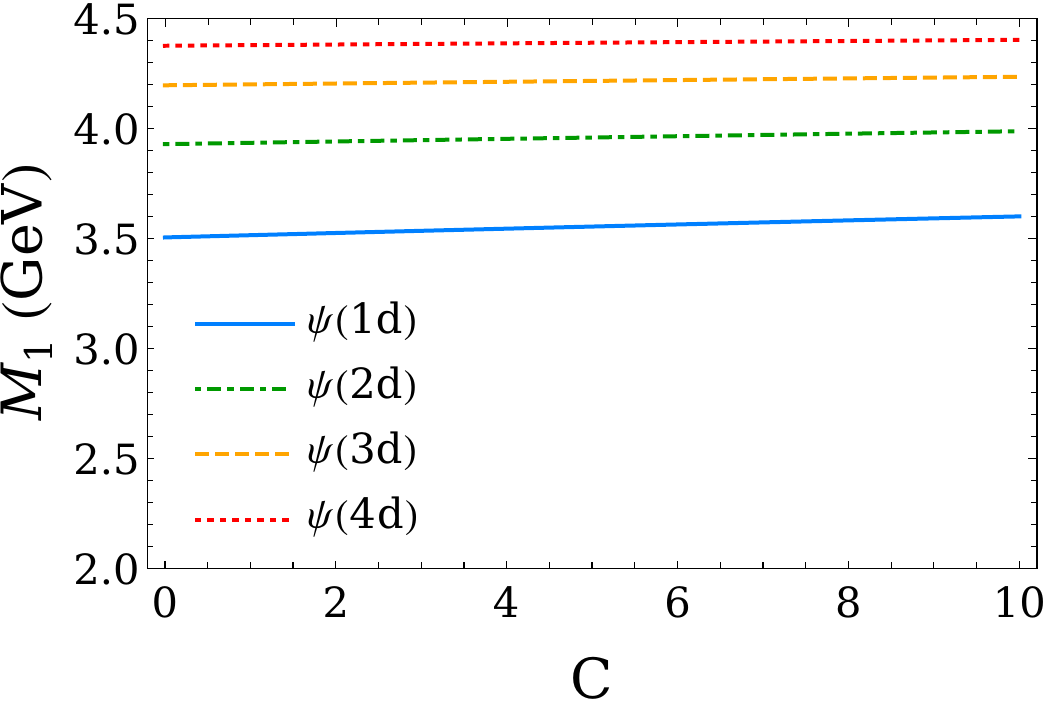}
    \includegraphics[scale=0.6]{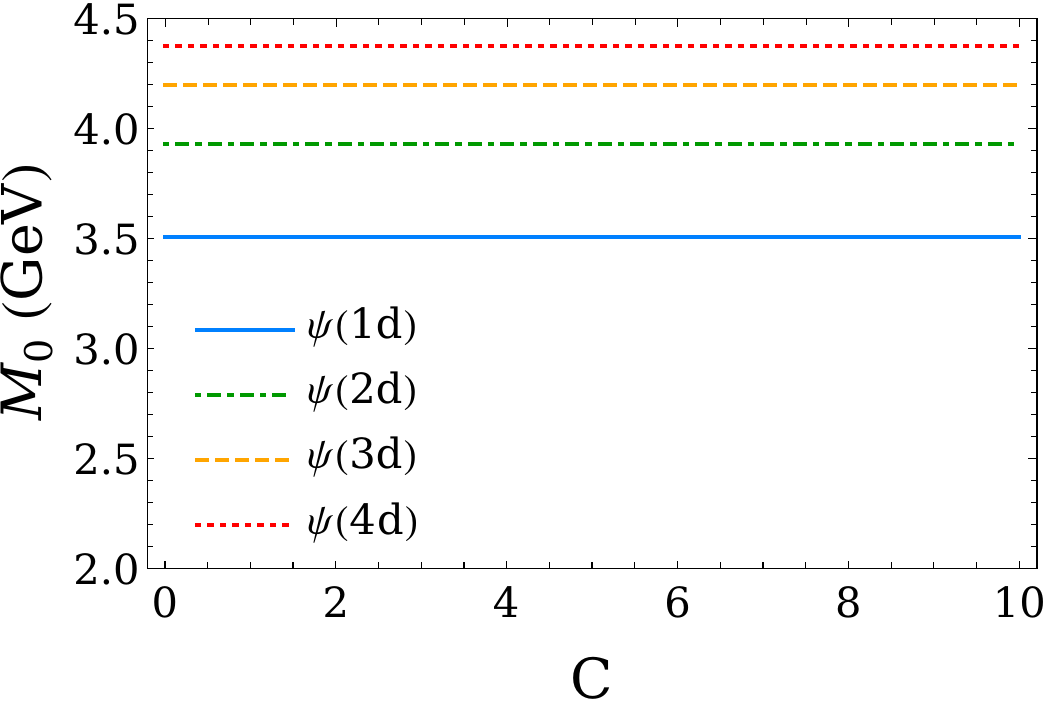}
    \includegraphics[scale=0.6]{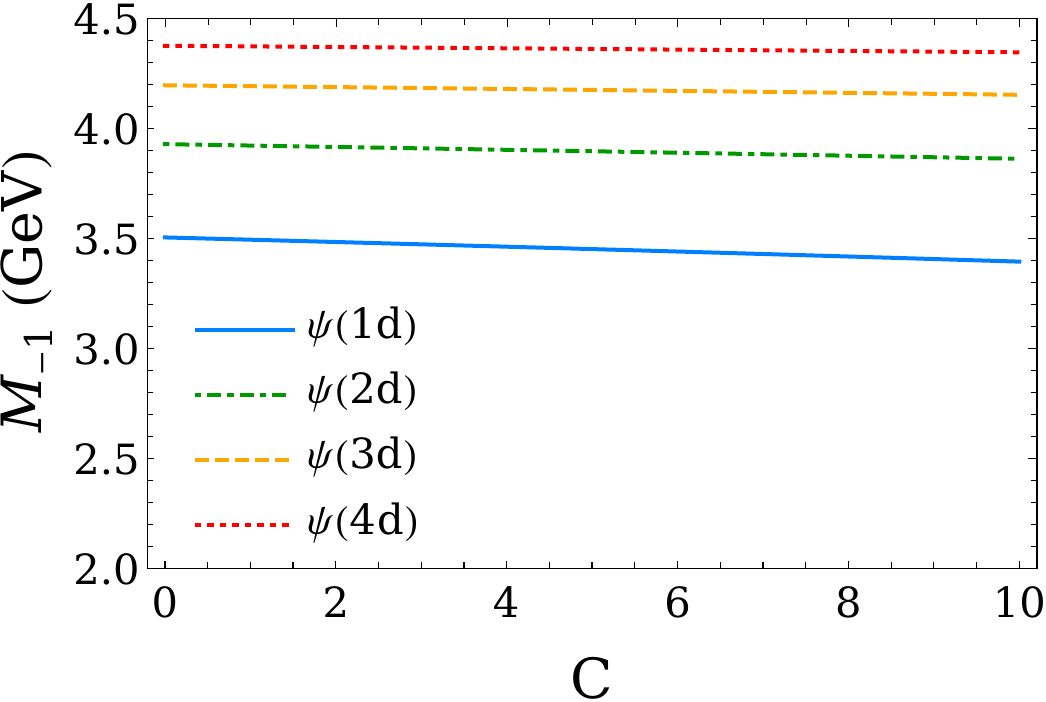}
        \caption{\textcolor{black}{Charmonium mass in relation to parameter $C$ for various energy and momentum levels (notation follows atomic physics quantum numbers) using $T = 168$ MeV}} % with quark charm mass $m_q = 1.275^{+0.025}_{-0.035}$ (GeV).}
    \label{massfigc}
\end{figure}

\begin{figure}[h]
    \includegraphics[scale=0.6]{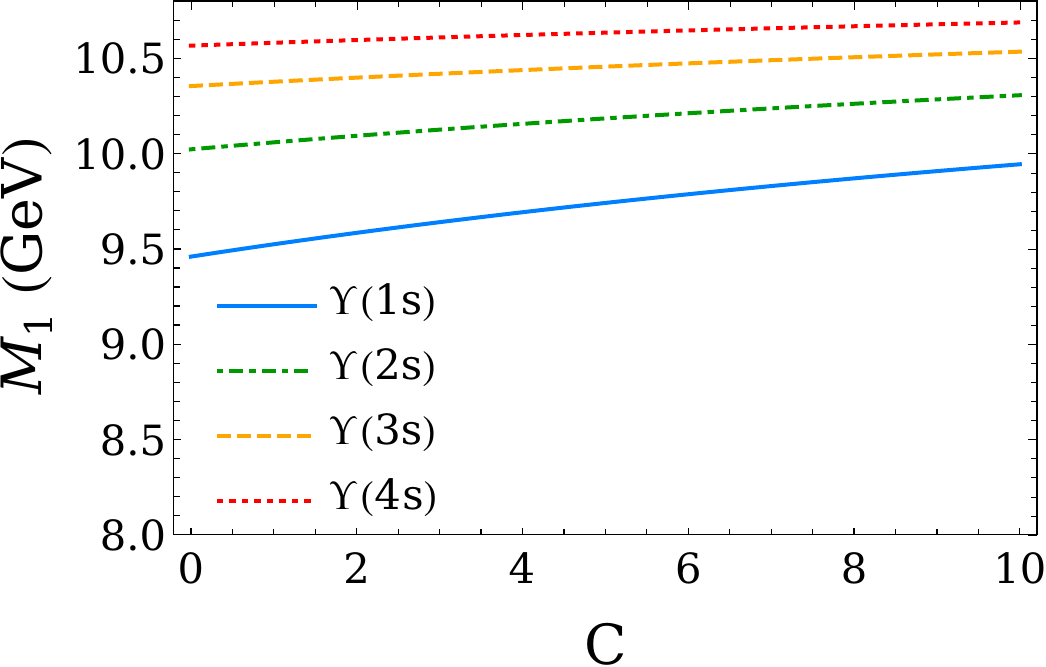}
    \includegraphics[scale=0.6]{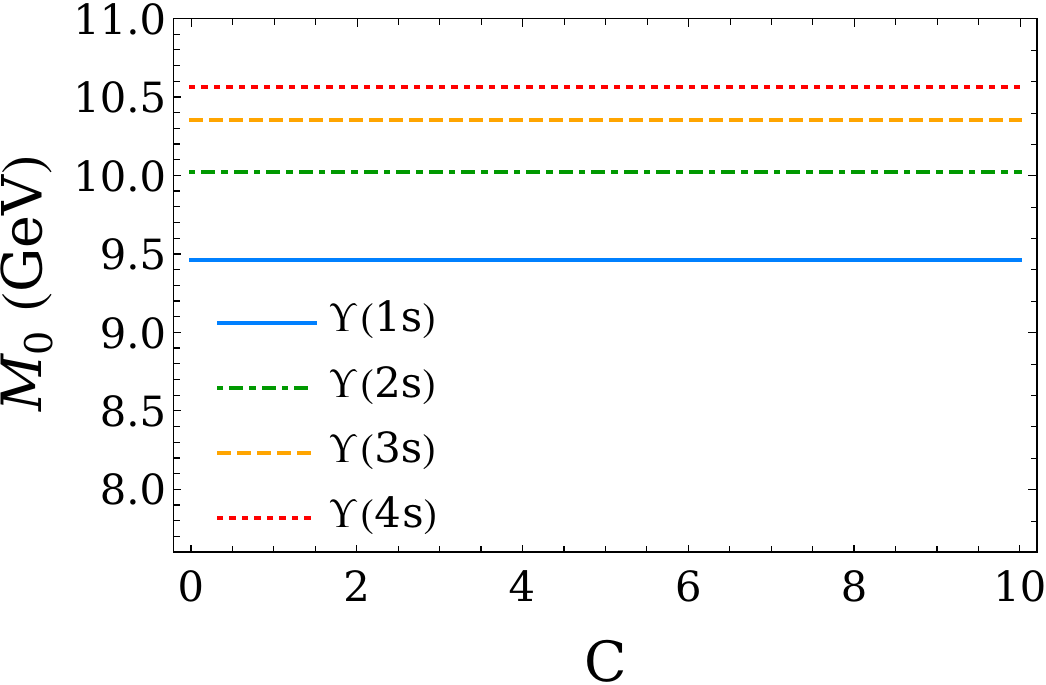}
    \includegraphics[scale=0.6]{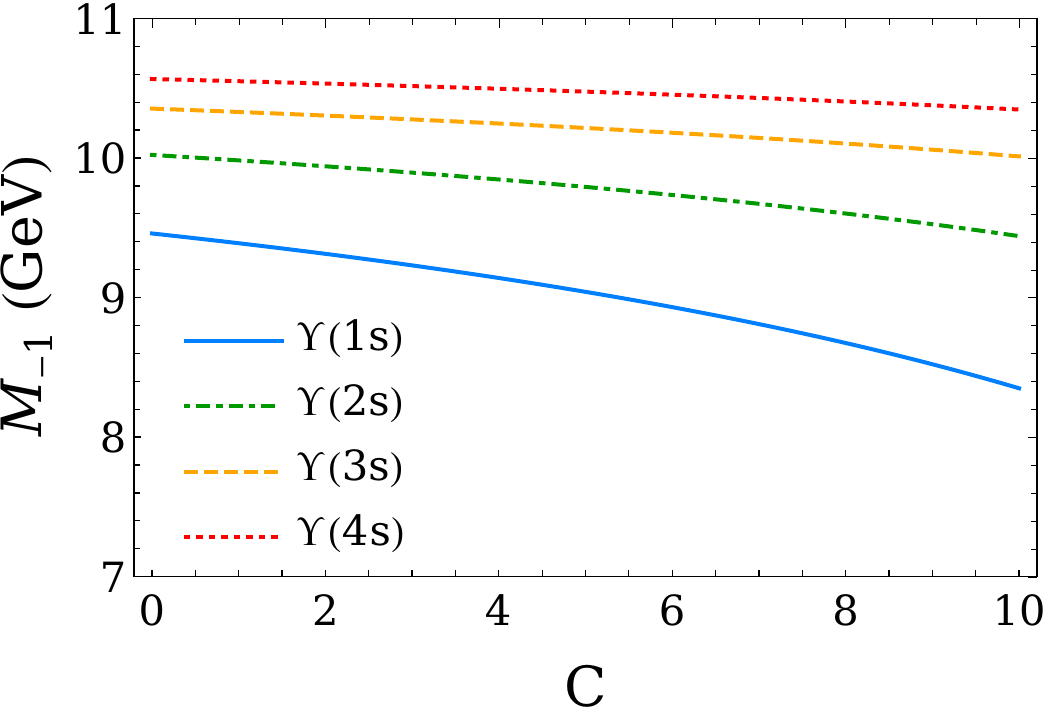}
    \includegraphics[scale=0.6]{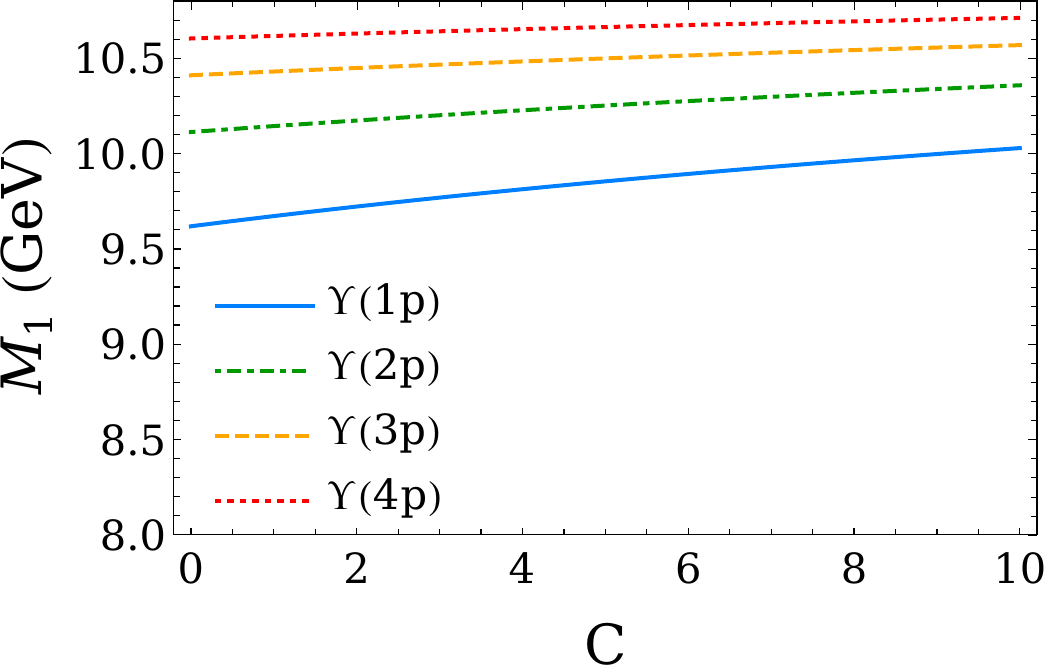}
    \includegraphics[scale=0.6]{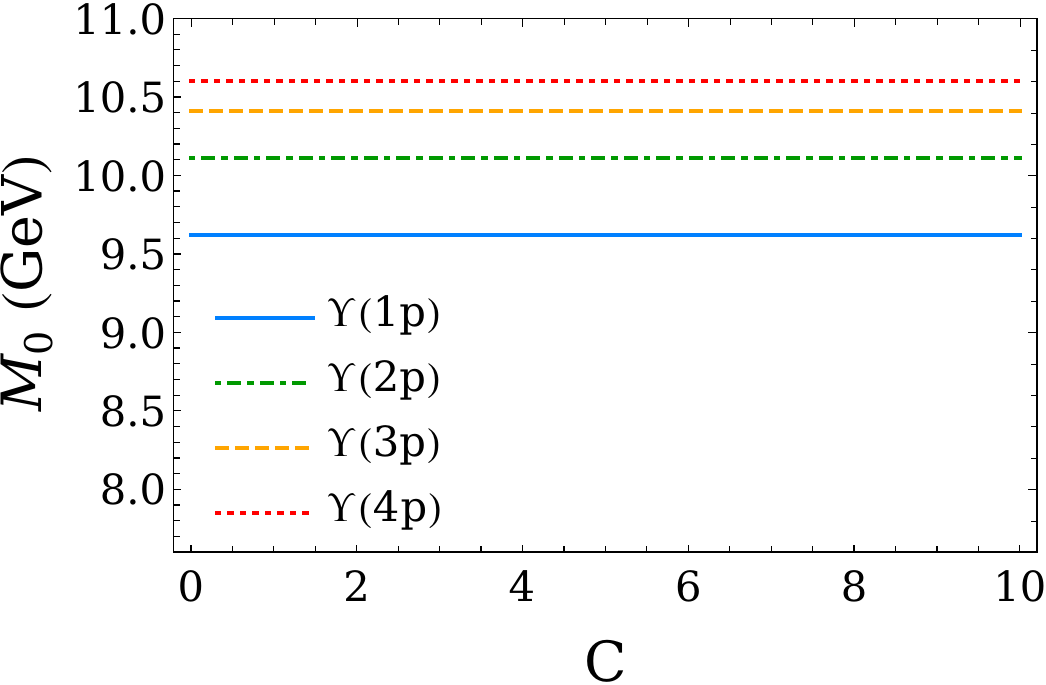}
    \includegraphics[scale=0.6]{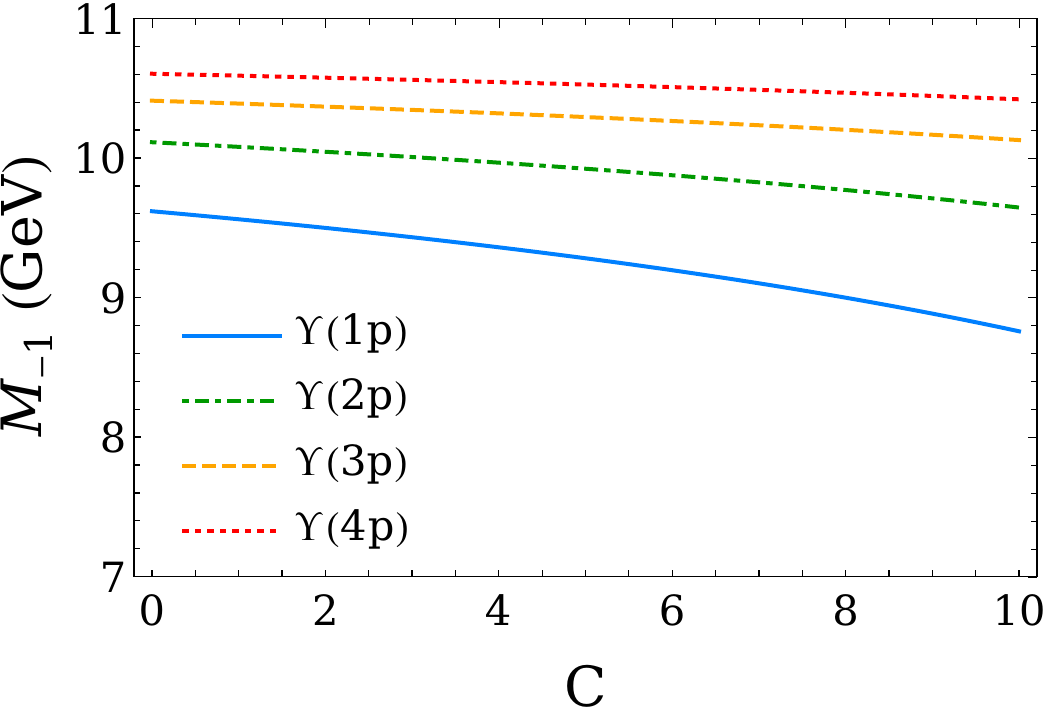}
    \includegraphics[scale=0.6]{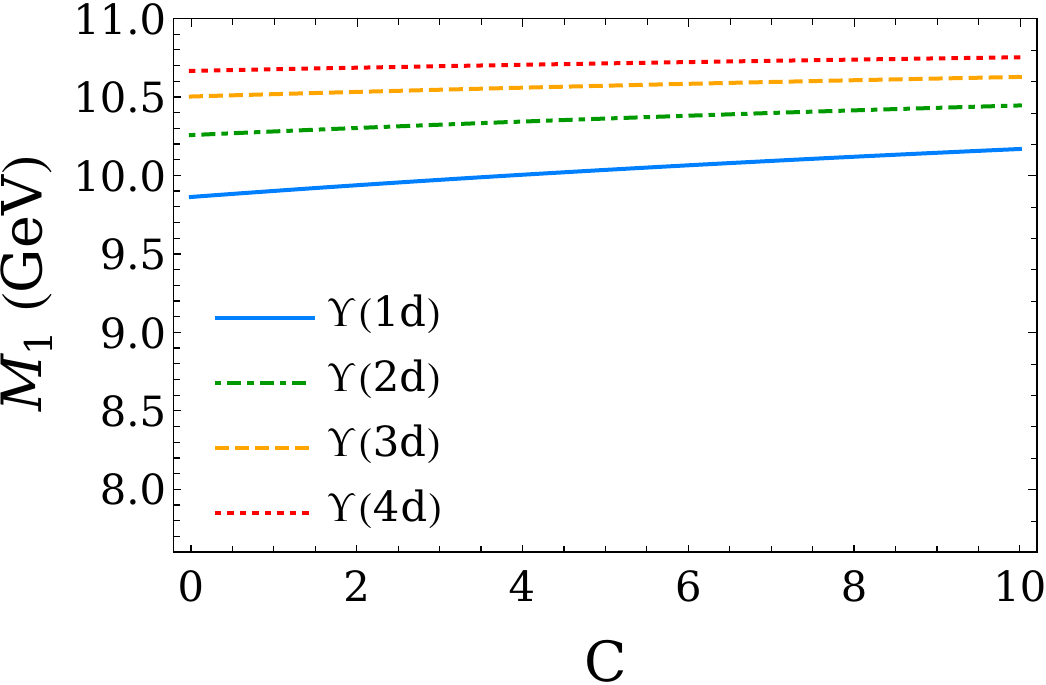}
    \includegraphics[scale=0.6]{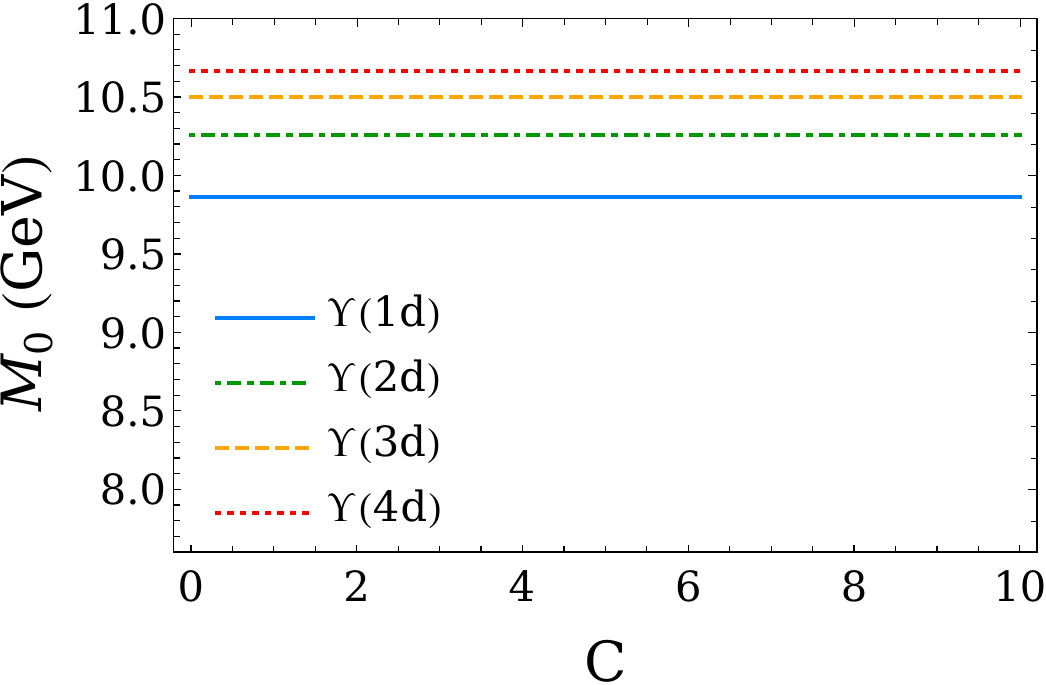}
    \includegraphics[scale=0.6]{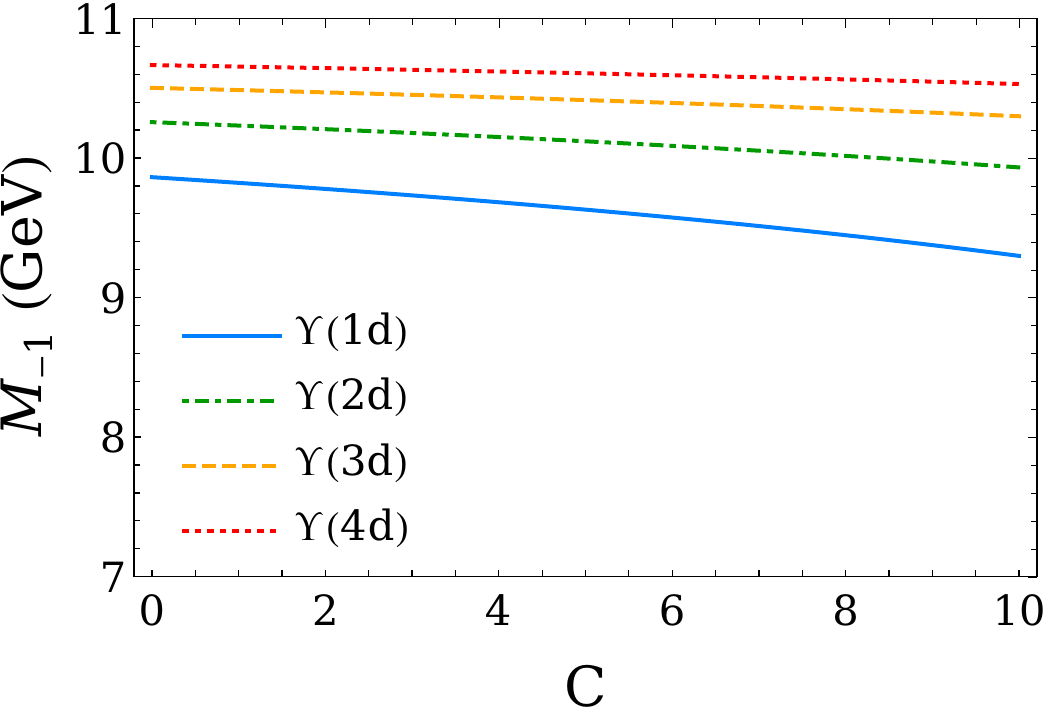}
        \caption{\textcolor{black}{Bottomonium mass in relation to parameter $C$ for various energy and momentum levels. Notation follows atomic physics quantum numbers using $T = 168$ MeV}} % with quark charm mass $m_q = 1.275^{+0.025}_{-0.035}$ (GeV).}
    \label{massfigb}
\end{figure}

\figref{massfigc} and \figref{massfigb} gives a quantitative value of the binding energy for quarkonia polarized with and opposite the vorticity as a function of the circulation.   Unsurprisingly, one is the opposite of the other.   This, however, is not experimentally detectable since spin alignment measurements do not distinguish between $m=\pm 1$ states.    The mass difference in \secref{massfigc}  would therefore appear as an impact parameter dependent {\em widening} of the quarkonium state.   However, since a similar widening occurs in any kind of in-medium interaction, particularly in interactions leading to the melting of quarkonium \cite{classic3}, a univocal proof of spin-orbit coupling can not be obtained by measuring $E_{\pm 1}$ alone.   As we will show in the next section \secref{offdiag}, however, off-diagonal matrix elements could be of help here.

\subsection{Vorticity and melting \label{dissoc}}
We can use the mass dependence calculated in the previous sub-section to 
 study, qualitatively, how the temperature for quarkonium melting (melting temperature, $T_{melt}$) changes under a non-inertial frame.   This calculation can be considered to be a semi-classical estimate of the imaginary part of the energy calculated in the previous subsection \secref{masssec}.

Using a semi-classical analysis with $\ave{p}\sim 1/r$ and $\ave{p^2}\sim 1/r^2$ we can write the energy by using
\begin{equation}
	E(r) = \frac{1}{2\mu r^2} -m_j\;\omega + V(r)
\end{equation}
For the \changed{$J/\psi$} melting, we shall use the potential that takes into consideration the Debye screen where $\lambda_D$ is Debye screening length, which is given by:
\begin{equation}
	V(r) = - \alpha_{eff}\frac{e^{-r/\lambda_D}}{r}
\end{equation}
Next, using the circulation theorem \ref{cir}. Then,
\begin{equation}
	E(r) = \left[\frac{1}{2\mu } - \frac{m_j\;C}{2\pi }\right] \frac{1}{r^2} - \frac{\alpha_{eff}e^{-r/\lambda_D}}{r}	
	\label{eneraux}
\end{equation}
 The bound state is defined when the energy \ref{eneraux} has a minimum, so we can write:

 \begin{equation}
	 \frac{dE(r)}{dr} = -\left[\frac{1}{\mu } - \frac{m_j\;C}{\pi }\right] \frac{1}{r^3} + \alpha_{eff}e^{-r/\lambda_D}\left[1+\frac{r}{\lambda_D}\right]\frac{1}{r^2} =0
 \end{equation}
,
then

\begin{equation}
	r\left[1+\frac{r}{\lambda_D}\right]\alpha_{eff}e^{-r/\lambda_D} = \frac{1}{\mu } - \frac{m_j\;C}{\pi }
	\label{dev}
\end{equation}

Making the following variable change $\tilde{r} = r/\lambda_{D}$, we get:

\begin{equation}
	f(\tilde{r})=\tilde{r}\left(1+\tilde{r}\right)e^{-\tilde{r}} = \frac{1}{\alpha_{eff}\lambda_D}\left[\frac{1}{\mu } - \frac{m_j\;C}{\pi }\right]
	\label{f}
\end{equation}

The maximum value of $f(\tilde{r})$ is $0.840$ at $\tilde{r} = 1.92$ in this limit we will have the non bound states. Thereby, we can write the inequality for this limit:
%The maximum value of $f(\tilde{r})$ is $f(C)_{max}$ in this limit we will have the non bound states. Thereby, we can write the inequality for this limit:

\begin{equation}
	\frac{1}{\alpha_{eff}\lambda_D}\left[\frac{1}{\mu } - \frac{m_j\;C}{\pi }\right] > 0.840 
\end{equation}

%To case where $\lambda_{D}\rightarrow \infty$ in \ref{dev} we obtain the charmonium radius in following form:

%\begin{equation}
%	r_{J/\Psi} = \left[\frac{1}{\mu } - \frac{m\;C}{\pi }\right]\left[\alpha_{eff}+\frac{m\;C}{2\pi}\right]^{-1}
%\end{equation}

\begin{figure}[h]
	    \includegraphics[scale=0.7]{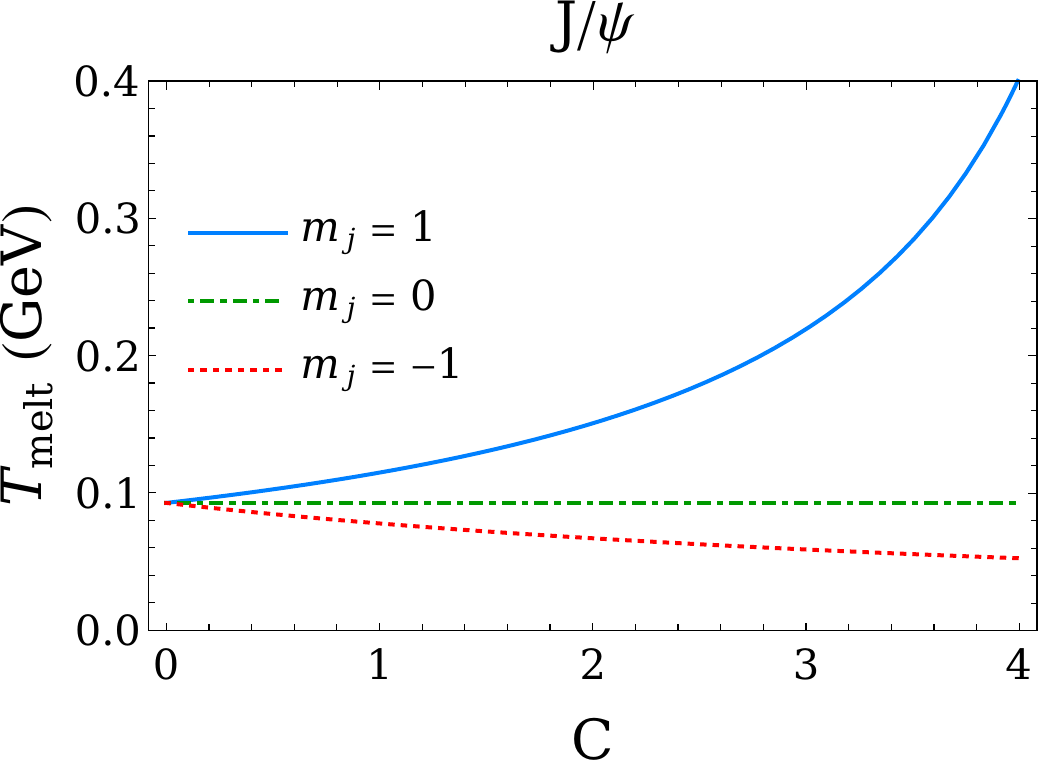}
	    \includegraphics[scale=0.7]{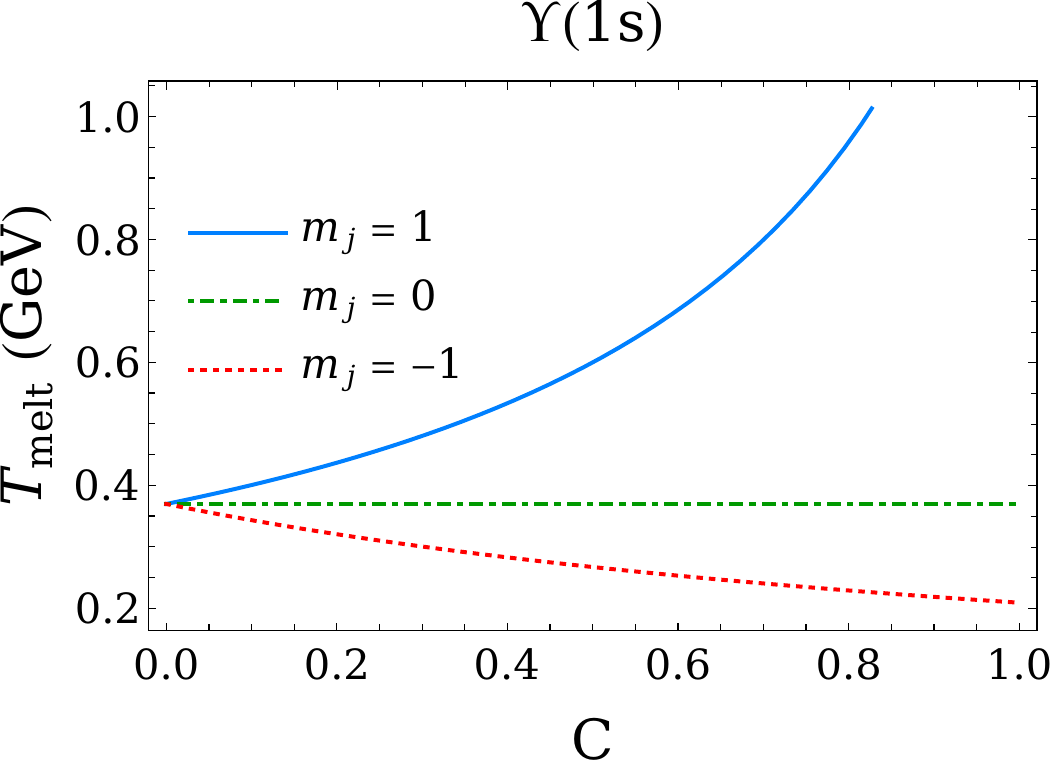}
    \caption{An estimate of melting temperature $T_{melt}$ for \changed{$J/\psi$} (left) and $\Upsilon$ (right) melting to different spin polarization $m$ values.}
    \label{tmeltfig}
\end{figure}

At this point, we will utilize the Debye lenght $\lambda_D$ value from the lowest-order perturbative QCD \cite{WONG} \footnote{The $\lambda_{D}$ value to non-inertial frames is not same as the inertial case. Nevertheless, the Debye mass to an inertial frame is an acceptable estimate because the non-inertial effect gives just a second-order contribution.}:

\begin{equation}
	\lambda_{D} = \sqrt{\frac{2}{9\pi\alpha_{eff}}}\frac{1}{T}  
\end{equation}
We can obtain an estimate for the melting temperature $T_{melt}$ in non-inertial frames using the equation \ref{f} in the $f(\tilde{r})$ maximum value, then:
\begin{equation}
	0.840\sqrt{\frac{2}{9\pi\alpha_{eff}}}\frac{1}{T_{melt}} = \frac{1}{\alpha_{eff}}\left[\frac{1}{\mu } - \frac{m_j\;C}{\pi }\right]
\end{equation}
So,
\begin{equation}
	T_{melt} = 0.840\sqrt{\frac{2\alpha_{eff}}{9\pi}}\left[\frac{1}{\mu } - \frac{m_j\;C}{\pi }\right]
	\label{tem1}
\end{equation}
\changed{We can see from figure \ref{tmeltfig} that the melting temperature $T_{melt}$ depends on the spin polarization quantum number $m$. In particular for $m = 1$ this temperature increases considerably.  For $m = 0$ it predictably does not change and for $m = -1$  it decreases but, and this is a fundamental point, it decreases a lot less than the increase for $m = 1$. Therefore,  in a vortical medium polarized quarkonia will be much more likely to survive, while un-polarized or anti-polarized quarkonium's survival probability does not change that much w.r.t. quarkonia in a non-rotating medium.}

\changed{This finding is qualitatively important since it shows that vorticity can link quarkonium suppression and polarization via "distillation".   In a vortical medium, the melting probability of quarkonium states will depend strongly on their polarization.      Such a mechanism will result, 
analogously to the $\eta$ \cite{eta}, to a strong and novel dependence of quarkonium abundance on centrality, which could be investigated quantitatively by a hydrodynamic model.
}

\changed{One can be "brave" and try to apply our potential model to 
the meson $\phi$, which is 
formed of strange and anti-strange quarks.
The strong-vorticity dependent melting might be able to explain the strong
spin alignment observed in experiment, which seems incompatible with Cooper-Frye freezeout \cite{star}.
In the picture described here, melting temperature for aligned $\phi$ rises a lot, while melting temperature for anti-aligned and non-aligned $\phi$
stays nearly the same.   Thus, the large apparent spin alignment of $\phi$s comes from a "distillation" process where only polarized $\phi$s survive, and this increases relative $\phi$ spin alignment.
Of course, applying potential models to $\phi$ is not justified theoretically
and requires model-building, so such a solution will need considerable quantitative and phenomenological development.
}

In the absence of a detailed hydrodynamic modeling of the correlation between charmonium suppression and polarization, however, distinguishing the effects outlined above from other variations of charmonium suppression with centrality, namely the effect of centrality dependence of temperature, looks complicated, with vorticity just adding an ``event-by-event widening'' to the processes of melting and regeneration of quarkonia.   
The next section, however, provides a direct experimentally measurable indication of spin-orbit non-equilibrium in quarkonium polarization measurements
\changed{\section{Density matrix elements and vorticity \label{offdiag}}}
Turning our attention back to the density matrix we can write this operator on basis of energy in the following way:

\begin{equation}
	\hat{\rho} = e^{-\beta\hat{H}} \eqcomma \beta = \frac{1}{T}
\end{equation}
Where the eigenvalues is given by the equation \ref{energy}. In this moment, we will make a rotation to lab frame, so:
\begin{equation}
	\hat{\rho}^r = U(\theta_r,\phi_r)\; \hat{\rho}\; U^{-1}(\theta_r,\phi_r)
    \label{rhorot}
\end{equation}

Now, we can expand the equation \ref{rhorot} in following way:

\begin{equation}
	\rho^r_{m,m'}=\sum_{m'', m'''} e^{i(m'''- m)\phi_r}d^j_{m,m''}(\theta_r)\rho_{m'',m'''}\left[d^j_{m''',m'}(\theta_r)\right]^{-1} \eqcomma \rho_{m'',m'''} = \frac{1}{Z} e^{\beta E_{m'''}}\delta_{m'',m'''}
	\label{densityMatrix}
\end{equation}

We can relate the density matrix coefficients from variable $C$ as both density matrix coefficients and energy variation depend on the parameter $C$. So we can relate these two values we obtain the figure \ref{den}. Then, we can relate the energy $(\Delta E_{m, m'} = E_m - E_{m'})$ with off-diagonal density matrix $Re[\rho_{-1,0}]$ and $Re[\rho_{-1,1}]$ using the values $\phi=0$, $\theta_r = 1.75\pm 0.10$ to Collins-Soper refer to transverse momentum range $2<p_T<4$ (GeV). 
\begin{figure}[h]
    \includegraphics[scale=0.7]{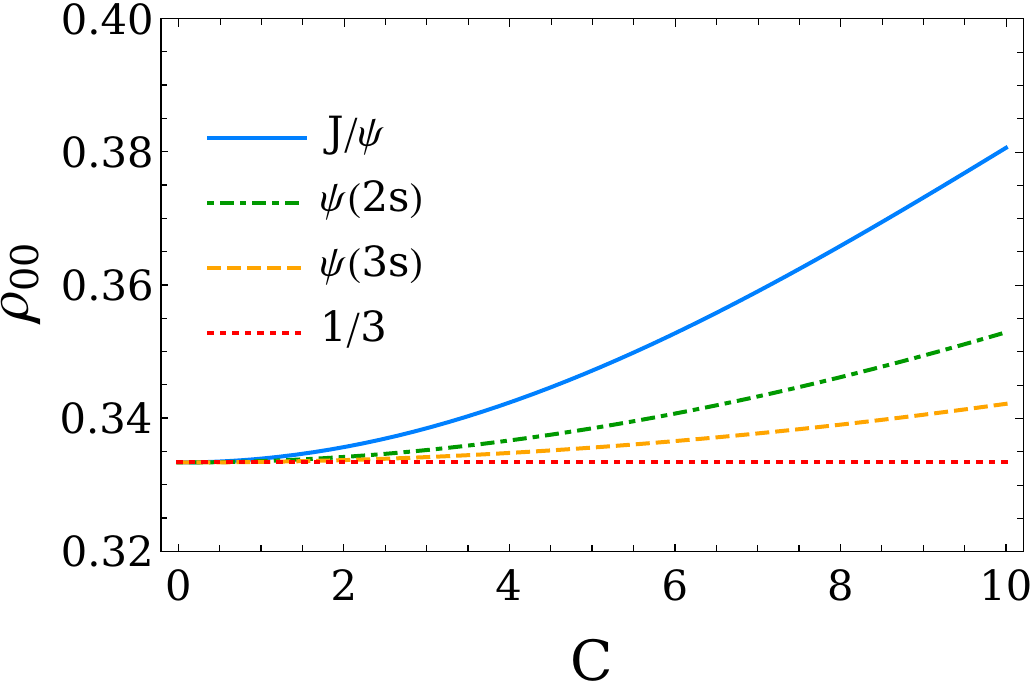}
    \includegraphics[scale=0.7]{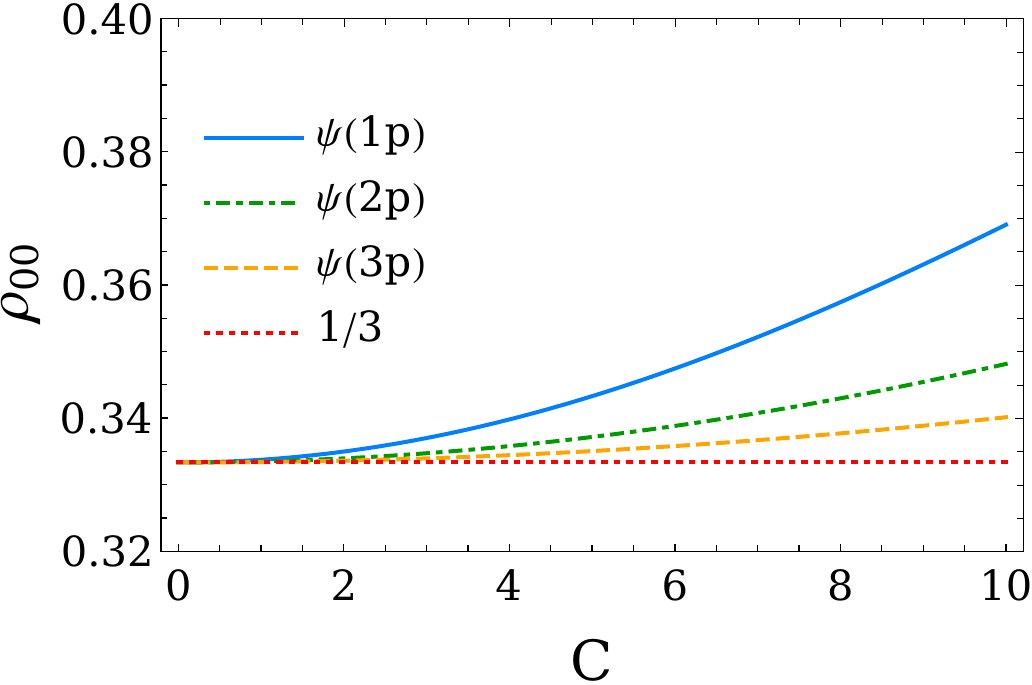}
    \includegraphics[scale=0.7]{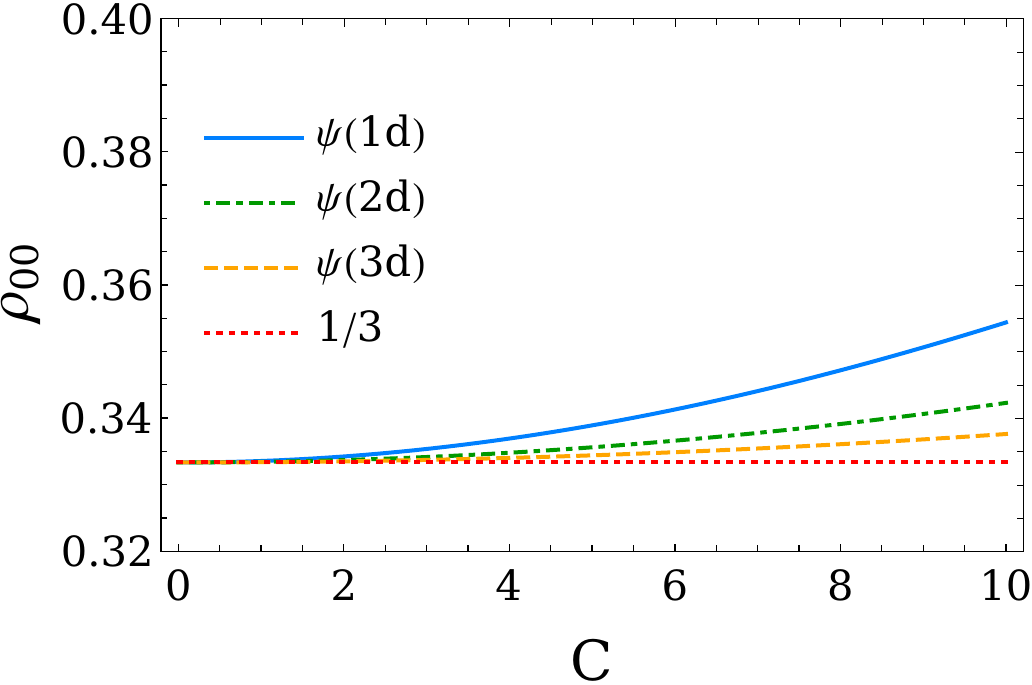}
    \includegraphics[scale=0.7]{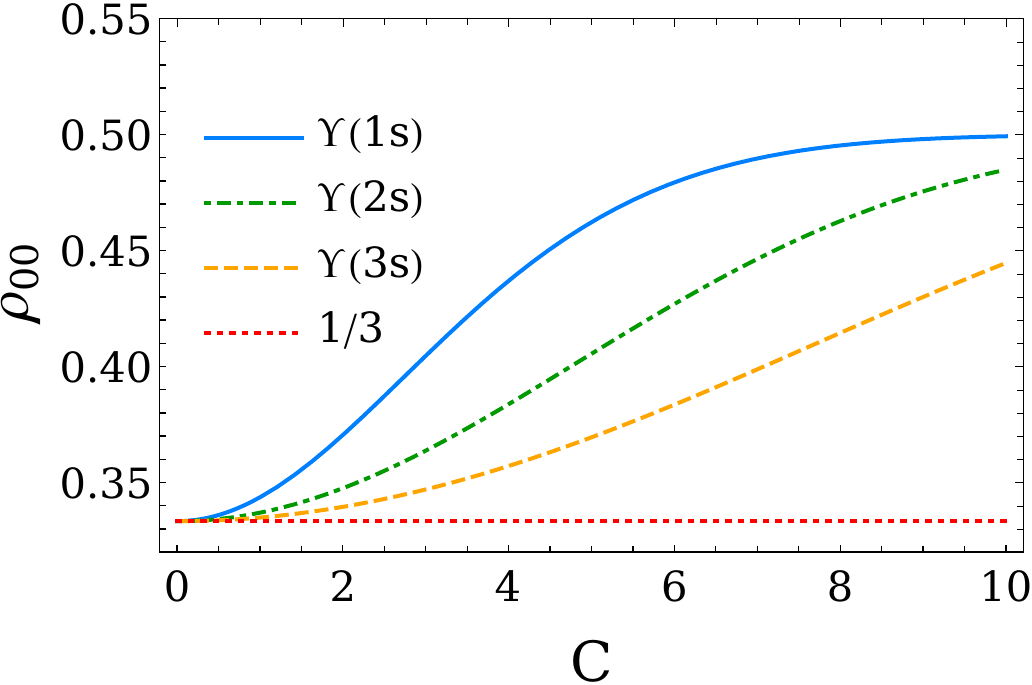}
    \includegraphics[scale=0.7]{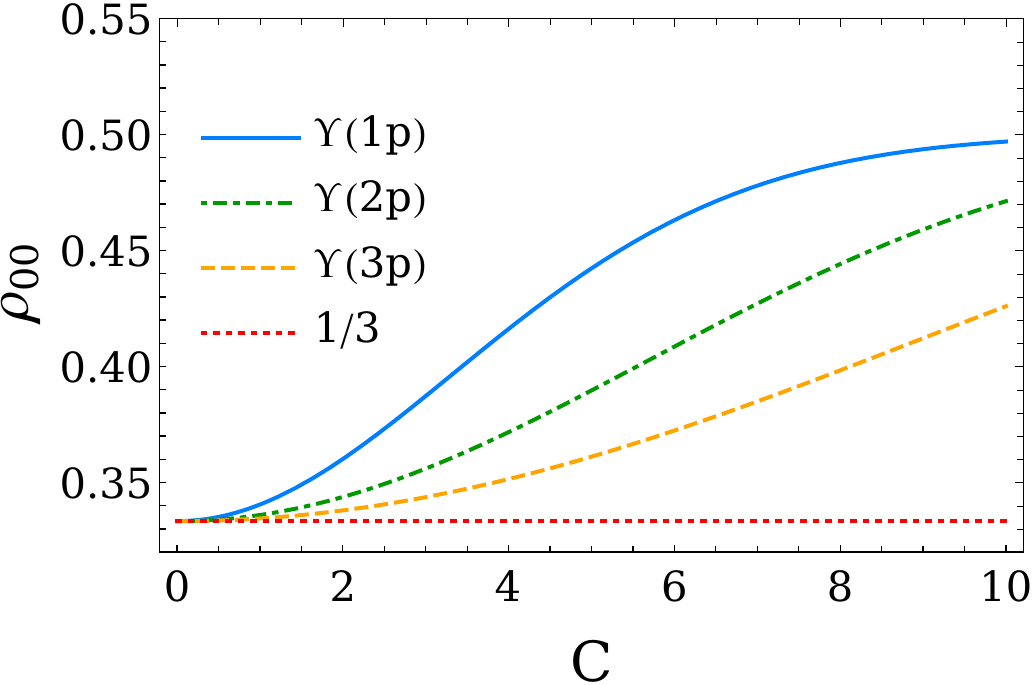}
    \includegraphics[scale=0.7]{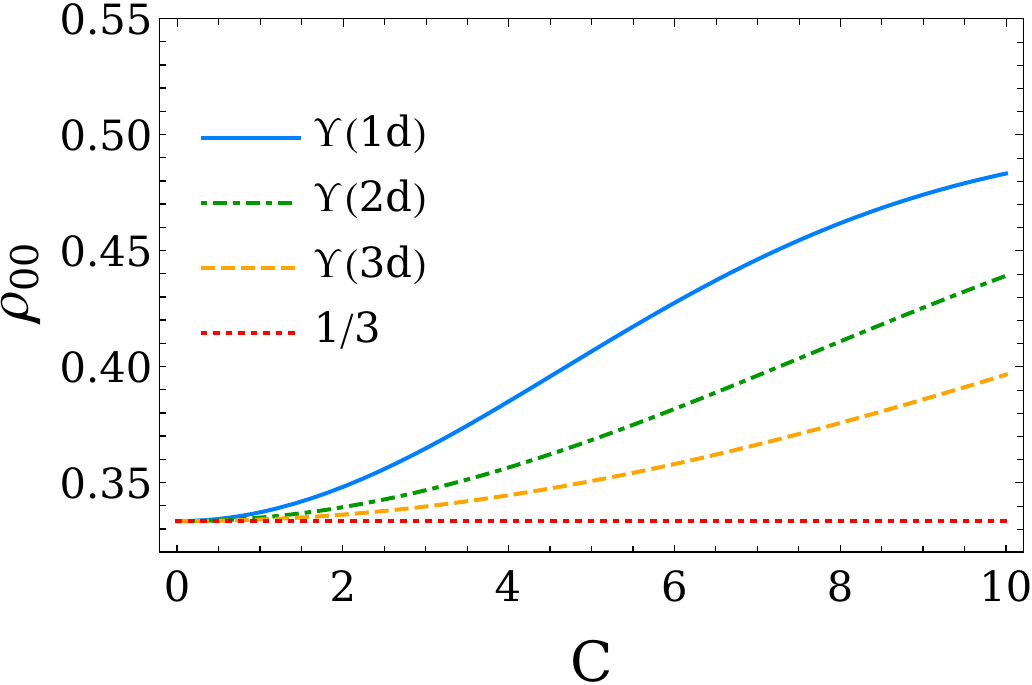}
        \caption{\textcolor{black}{The alignment factor $\rho_{00}$ in relation C for charmonium and bottomonium using $T = 168$ MeV.}}
    \label{den2}
\end{figure}

\begin{figure}[h]
    \includegraphics[scale=0.7]{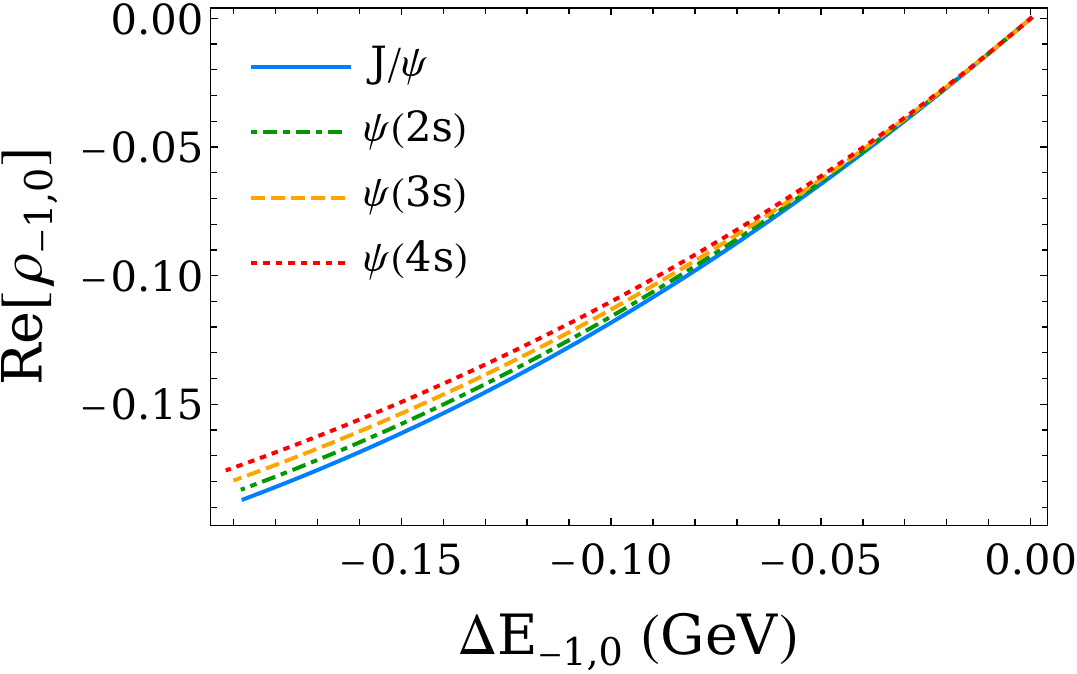}
    \includegraphics[scale=0.7]{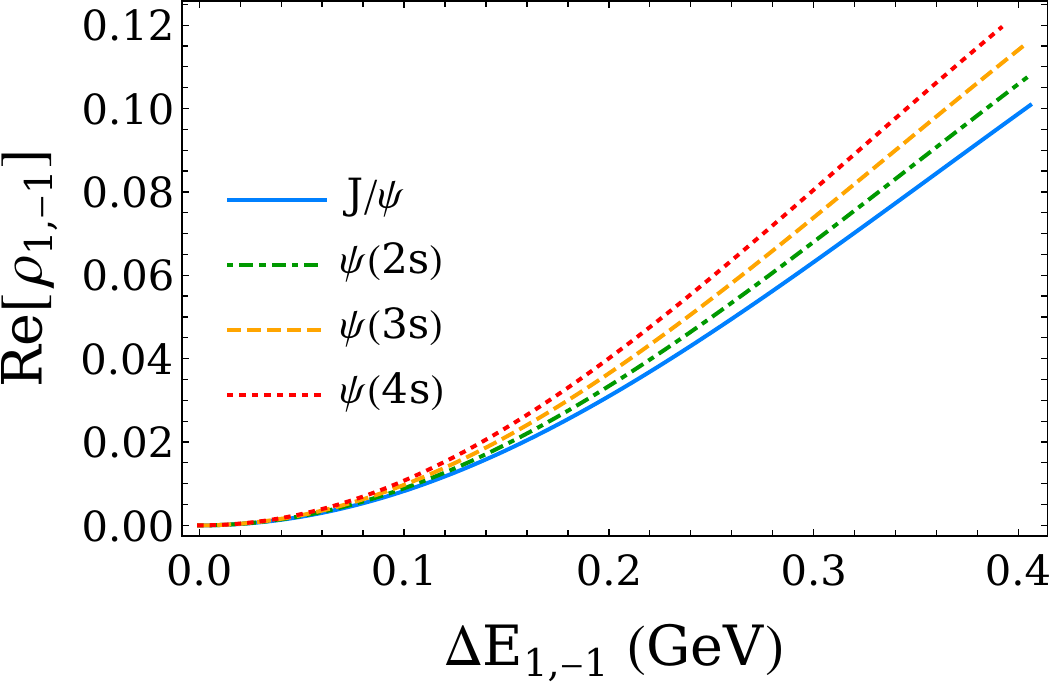}
    \includegraphics[scale=0.7]{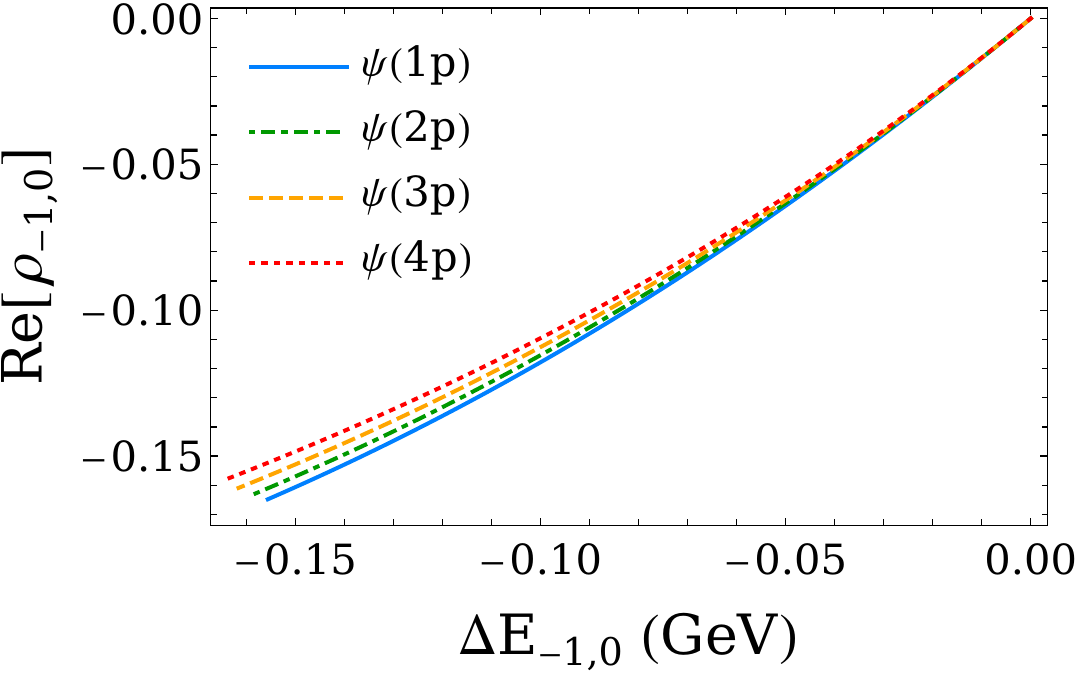}
    \includegraphics[scale=0.7]{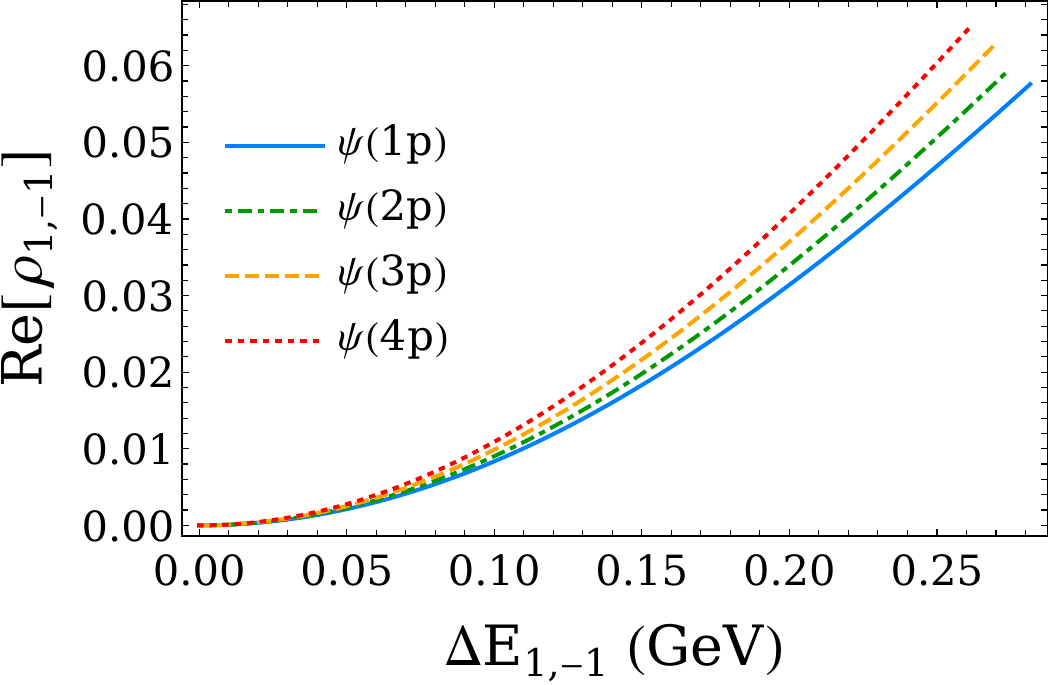}
    \includegraphics[scale=0.7]{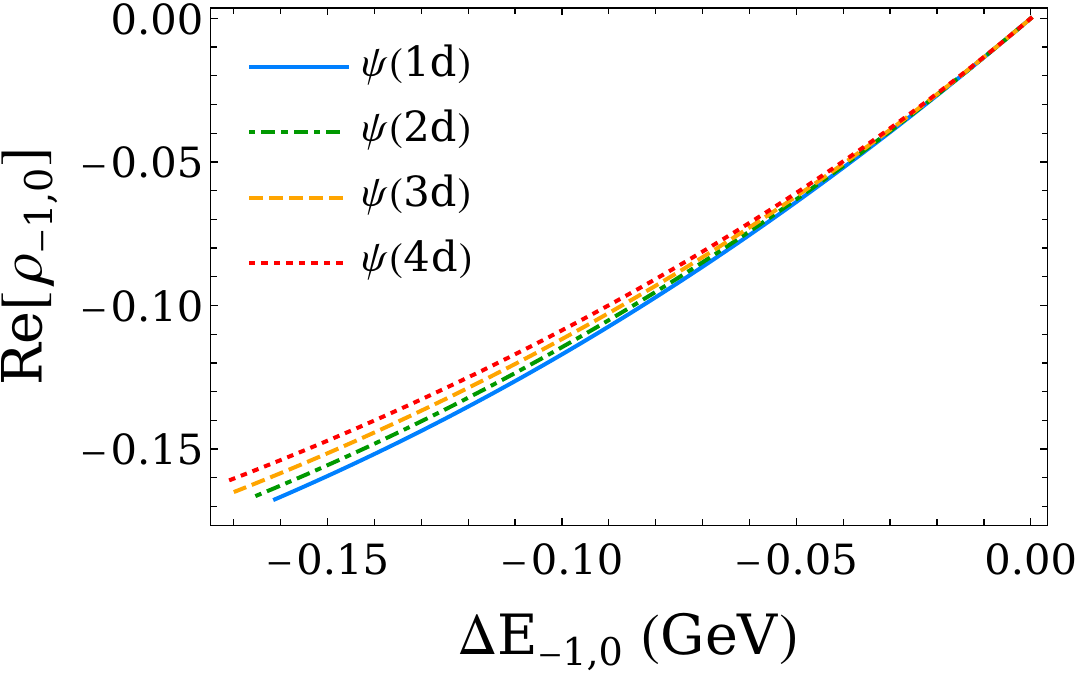}
    \includegraphics[scale=0.7]{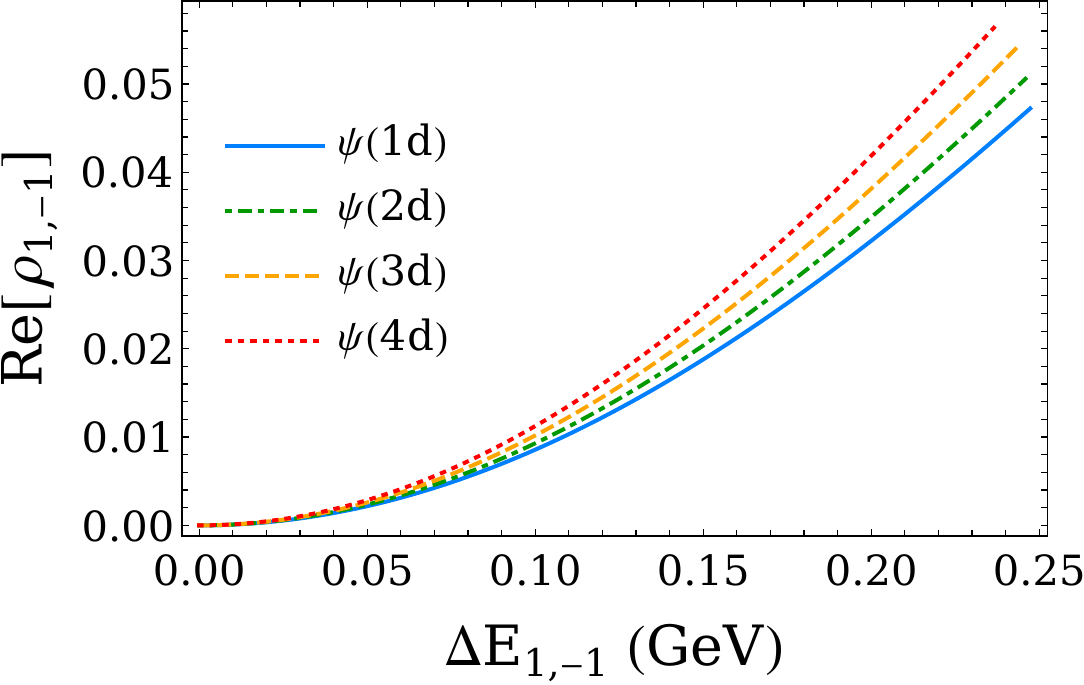}
    \caption{\textcolor{black}{The off-diagonal density matrix components in relation to energy variation $\Delta E_{m,m'}$ using $T = 168$ MeV.}}
    \label{den}
\end{figure}

\begin{table}[!b]
\changed{
% \begin{tabular}{|p{2cm}|p{2cm}|p{3cm}|p{2cm}|p{2cm}|p{3cm}|}
% \begin{tabular}{p{2cm}p{2cm}p{3cm}p{2cm}p{2cm}p{3cm}}
\begin{tabular}{|c|c|c|c|c|c|c|}
\cline{2-7}
	\multicolumn{1}{c|}{}     & $m \ (GeV)$ & $\alpha_{eff}$ & $b \ (GeV^2)$  & $\delta \ (GeV)$  & $\mu \ (GeV)$ &$\theta_r$\ (rad) \\\hline
	$c\bar{c}$  &  1.209      &  1.244         &  0.2           &  0.231            &   0.6045                    & $1.75 \pm 0.10$  \\\hline
	$b\bar{b}$  &  4.823      &  1.569         &  0.2           &  0.378            &   2.4115                    & $1.58 \pm 0.10$ \\\hline
\end{tabular}
	\caption{\label{tab_quarkonium2} \textcolor{black}{Eigenvalue quarkonium parameters obtained from \cite{AHM} with $\theta_r$ determined in system equation \ref{phirdef} to Collins-Soper frame to $J/\psi$ and $\Upsilon$ with respectively ranges $4 < p_T < 6\ (GeV)$ and $p_T < 15\ (GeV)$.}}
}
\end{table}
As can be seen in \figref{den2} the obtained $\rho_{00}$ are compatible with the estimates shown in \secref{prelim}

    % \changed{We can obtain the same off-diagonal density matrix components in relation to energy to charmonium using the parameters $\alpha_{eff} = 1.244$, $b = 0.2$, $\delta = 0.231$ and $\mu_c = m_c/2 $ with $m_c = 1.209$ GeV  we can see in figure \ref{den}. We can obtain the same off-diagonal density matrix components in relation to energy to botomonium using the parameters $\alpha_{eff} = 1.569$, $b = 0.2$, $\delta = 0.378$ and $\mu_b = m_b/2 $ with $m_b = 4.823$ GeV we can see in figure \ref{bot}\cite{AHM}.  }

\begin{figure}[h]
    \includegraphics[scale=0.7]{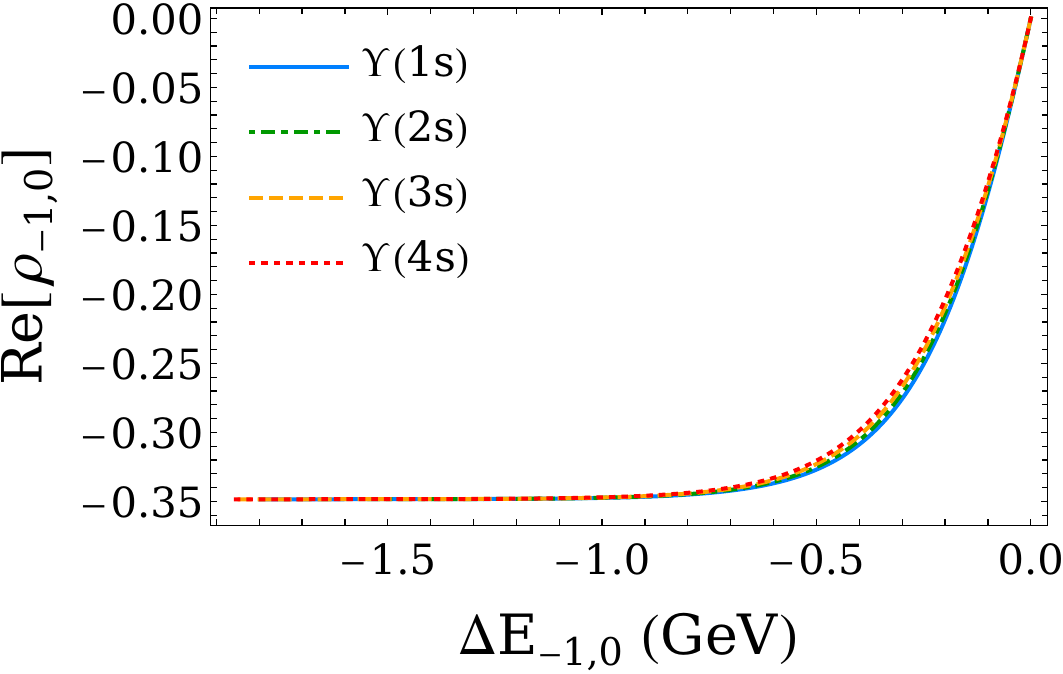}
    \includegraphics[scale=0.7]{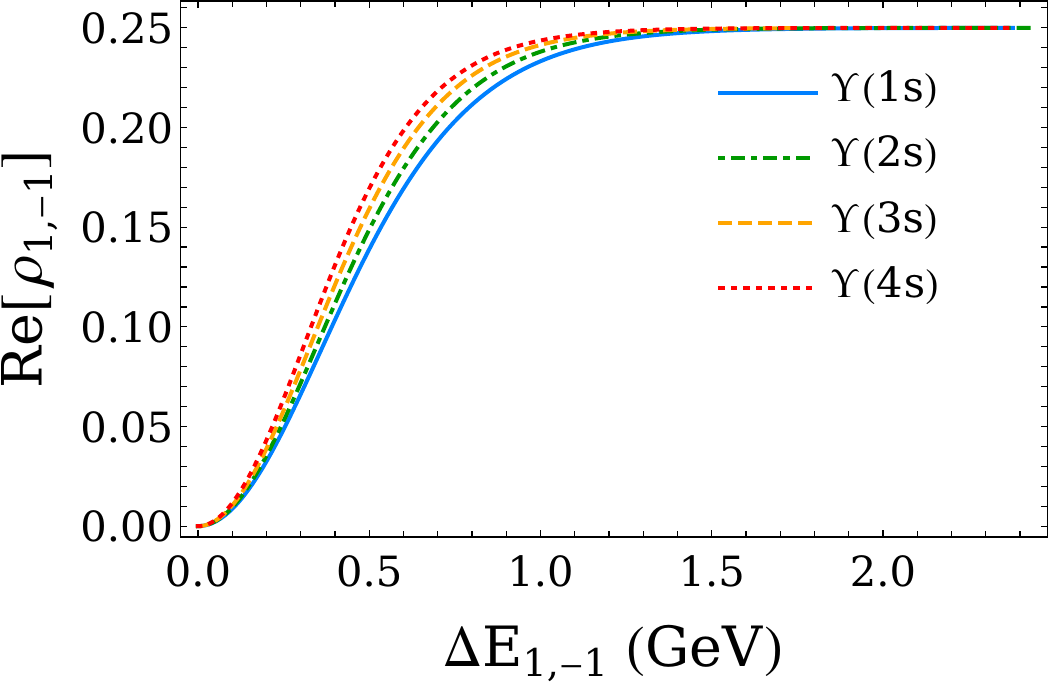}
    \includegraphics[scale=0.7]{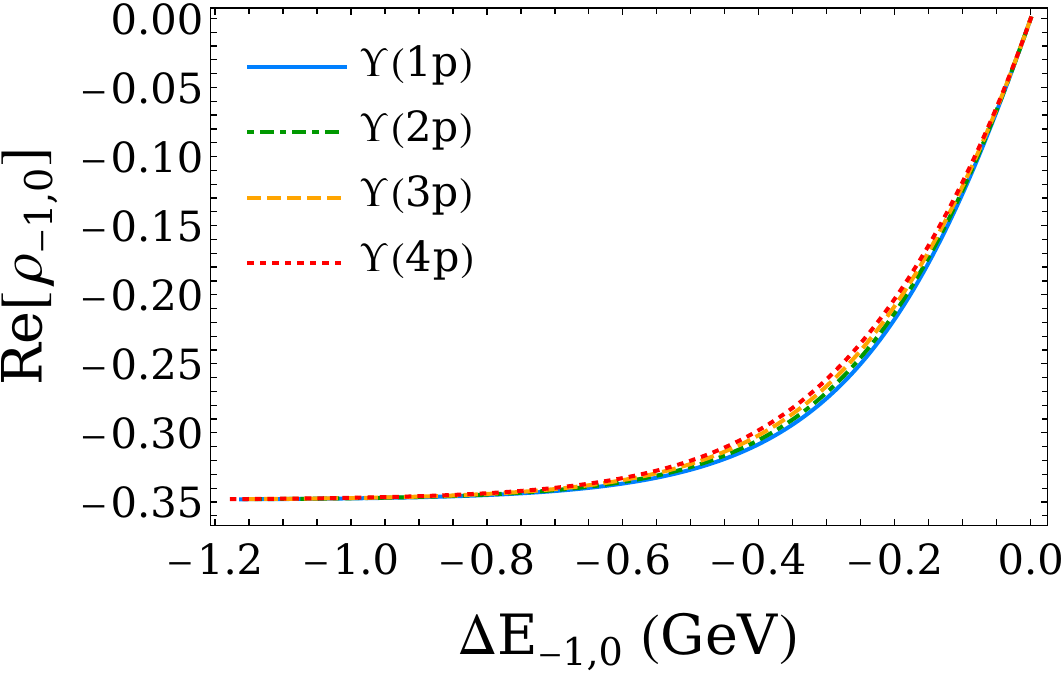}
    \includegraphics[scale=0.7]{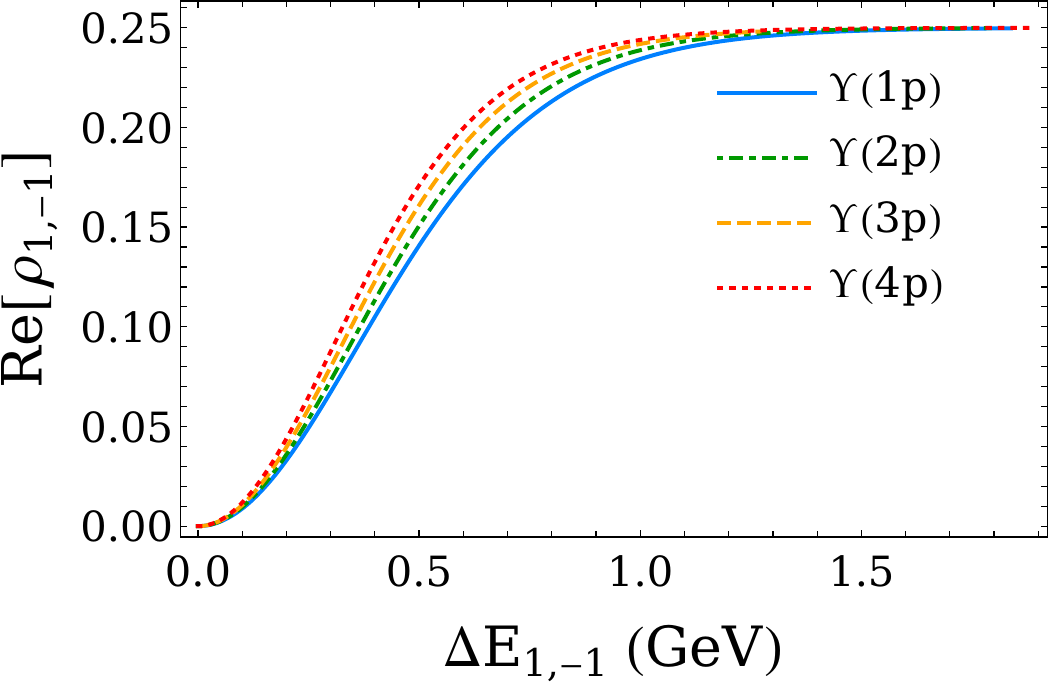}
    \includegraphics[scale=0.7]{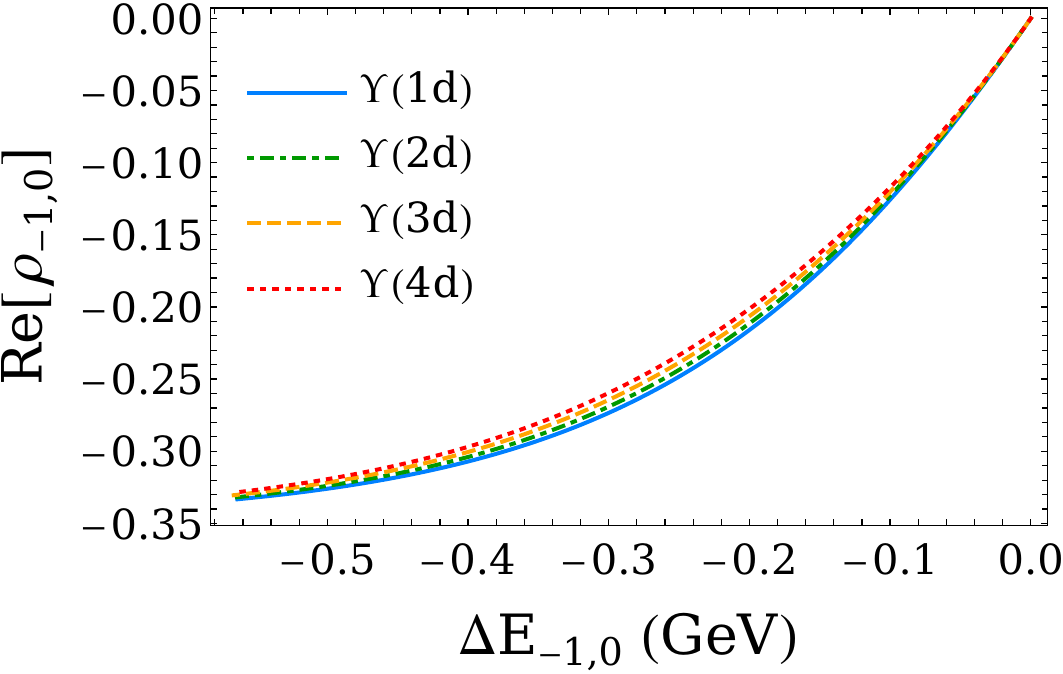}
    \includegraphics[scale=0.7]{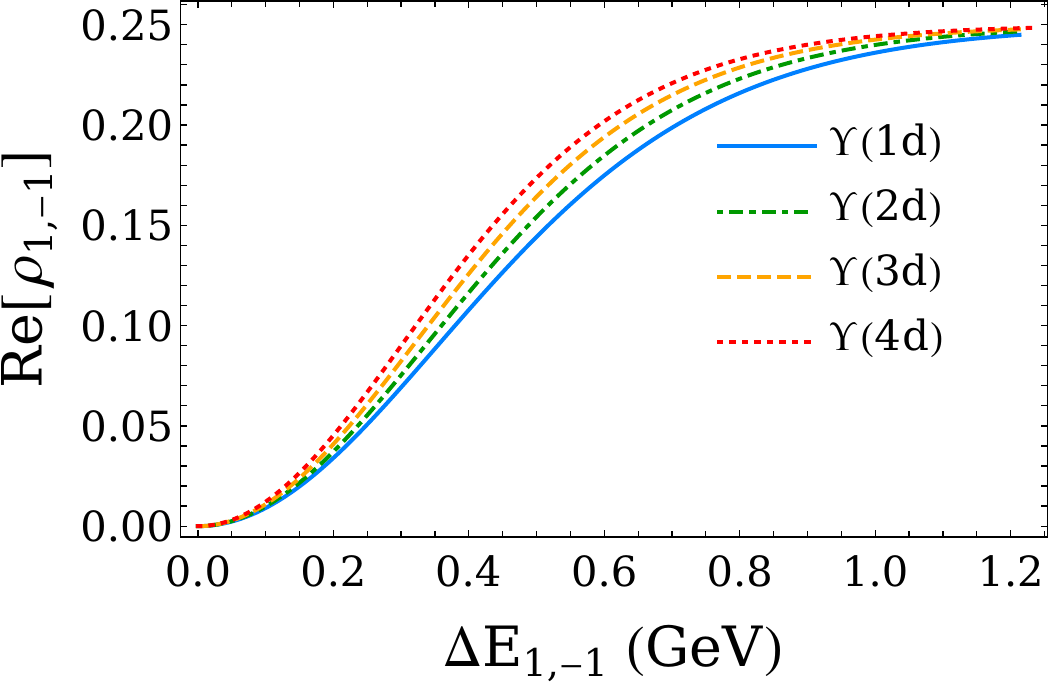}
    \caption{\textcolor{black}{The off-diagonal density matrix components in relation to energy variation $\Delta E_{m,m'}$ using $T = 168$ MeV.}}
    \label{bot}
\end{figure}

\changed{We can obtain the off-diagonal density matrix components in relation to the quarkonium energy using the parameters shown in Table \ref{tab_quarkonium2}. In figure \ref{den} and \ref{bot}, we can note the relation between the alignment factor,  $\rho_{00}$, with the circulation parameter C. It is evident that an increase of C increments $\rho_{00}$; however, this increase will depend on the type of meson. As we can see the bottomonium alignment factor is larger than the charmonium when compared under the same parameter C.} 

Unlike the figures in the previous sections, the $x,y$ axes of the plots in this section are independently measurable.   $\Delta E$ will be manifest event-by-event as an invariant mass correlation.   $\rho_{i,j}$ can be obtained \cite{wang} from $dN/d\Omega$ according to a harmonic analysis of Eq. \eqref{coefftab}.   Barring acceptance effects, there is no non-vortical dynamics capable of generating a correlation between these two quantities. \textcolor{black}{The most direct analysis to do is a correlation between $\lambda_{\theta,\phi,\theta \phi}$ and the invariant mass and width.  A definite signal of the correlation can be used, using the functions described in this section, to extract the angle between vorticity and heavy quark spin polarization.  This can then be used to constrain models of spin hydrodynamics such as \cite{uscausality,uscas2,usdiss,relax,jeon,gursoy,hongo,hongo2}.} This section's results, therefore, can be used as a baseline for developing an experimental analysis capable of probing non-equilibrium between spin and vorticity using quarkonium probes.

The most direct analysis to do is a correlation
 between $\lambda_{\theta,\phi,\theta \phi}$ and the invariant mass and
 width.  A definite signal of the correlation can be used, using the
 functions described in this section, to extract the angle between
 vorticity and heavy quark spin polarization.  This can then be used to
 constrain models of spin hydrodynamics such as
 \cite{uscausality,uscas2,usdiss,relax,jeon,gursoy,hongo,hongo2}. This
 section's results, therefore, can be used as a baseline for developing
 an experimental analysis capable of probing non-equilibrium between
 spin and vorticity using quarkonium probes.
\section{Conclusion}
The results obtained in this work are certainly to be considered more like a rough estimate than a quantitative analysis, since they use an early version of the Cornell potential model.   However we think that this model is good enough to get a physical intuition of the problem of linking spin-vorticity non-equilibrium to the quarkonium state in a rotating frame, and, respectively, the quarkonium state to experimental observables.  As section \ref{potential} illustrates, our formalism in fact reproduces the reasonable expectation of how the binding energy (mass) and melting probability (width) of the quarkonium state respond to rotation.

Section \ref{offdiag} therefore takes the model further and examines what happens when rotation and spin are not aligned, and suggests experimental measurements.
A positive experimental observation of the observables described in the previous section will provide evidence that quarkonium state polarization is not thermally aligned, but in fact ``remembers'' a polarization state which is quite distinct from the vorticity state.

As argued in \secref{prelim} and \figref{translong}, one in fact expects quarks to be aligned transversely, which would also follow from initial state dynamics \cite{pqcd1,pqcd2} and from the large transverse vorticity present in the beginning of the collision \cite{bec}.   However, the vortex where the quarkonia will be could up could well be aligned to the flow gradients that developed on the time-scale of a hydrodynamic expansion, which will be in the longitudinal direction.    The combination of the two will result in a density matrix exhibiting large corrections to the expectation value in \cite{zanna}, with the magnitude of these correcting correlated to the invariant mass of quarkonium.

A very interesting associated result is the strong dependence, non-monotonic in both $T$ and $m$, of the dissociation temperature $T_{melt}$ of quarkonium shown in \secref{dissoc}.   It suggests a new "distillation" mechanism for quarkonium polarization, where aligned quarkonia have a stronger probability to survive in medium while non-aligned and anti-aligned quarkonia simply melt.     Speculatively considering the $\phi$ a quarkonium state could explain the strong alignment signal seen in \cite{star}, though some phenomenological work is needed to confirm such a model is viable.

So far, as seen in \secref{prelim} there is no direct experimental evidence for such non-equilibrium.  In the short term, an azimuthal modulation of $\lambda_{\theta,\phi,\theta\phi}$ (\figref{pl} ) could provide a strong indication that such non-equilibrium should be investigated.  In the long term, 
an experiment with sufficient resolution in invariant mass and capable of reconstructing the spin alignment could be able to perform the analysis advocated in \secref{offdiag}, providing more direct experimental evidence.

Quarkonium polarization, therefore, could well be a very promising observable for experimentally probing spin hydrodynamics in the quark-gluon plasma, a theoretically interesting and challenging topic that so far had little contact with phenomenology.

GT thanks CNPQ bolsa de produtividade 306152/2020-7, bolsa FAPESP 2021/01700-2 and participation 
in tematic FAPESP, 2017/05685-2 and grant BPN/ULM/2021/1/00039 from the Polish National Agency for Academic
Exchange
K.J.G. is supported by CAPES doctoral fellowship
88887.464061/2019-00.  This work was supported in part by the Polish National Science Centre Grant Nos 2018/30/E/ST2/00432. 
We thank the hospitality of the Jagellonian university when part of this work was performed.
\appendix
\section{Nikiforov-Uvarov method\label{method}}
We will use the method developed by Nikiforov and Uvarov to solve a differential equation where it is possible to reduce to hypergeometric function. We can do it too to Schr\"{o}dinger equation, which can write in the following form:
\begin{equation}
	\psi^{''}(s) + \frac{\tilde{\tau}(s)}{\sigma(s)}\psi^{'}(s) + \frac{\tilde{\sigma}(s)}{\sigma^2(s)}\psi(s)=0 
	\label{hp}
\end{equation}
\changed{Using the variable separation technique, 
\begin{equation}
        \psi(s) = \phi(s)y(s)
        \label{trans}
\end{equation}
}
Using the variable separation technique, we can see that the equation must satisfy the following expression \cite{NU}:
\begin{equation}
	\frac{\phi^{'}(s)}{\phi(s)} = \frac{\pi(s)}{\sigma(s)}
	\label{phi}
\end{equation}
so we can simplify as
\begin{equation}
	\sigma(s)y^{''}(s) +\tau(s)y(s) + \lambda y(s) = 0
\end{equation}
At this moment, we need to define the $\pi(s)$ function in the following form:
\begin{equation}
	\pi(s) = \frac{\sigma^{'}(s)-\tilde{\tau}(s)}{2}\pm \sqrt{\left(\frac{\sigma^{'}(s)-\tilde{\tau}(s)}{2}\right)^2 - \tilde{\sigma}(s)+k\sigma(s)}
	\label{pi}
\end{equation}

and $\lambda$ parameter is given by:

\begin{equation}
	\lambda = k + \pi^{'}(s)
	\label{lambda1}
\end{equation}

The Schr\"{o}dinger equation eigenvalue obtained from this method is given by:

\begin{equation}
	\lambda_n = -n\tau^{'}(s) -\frac{n(n-1)}{2}\sigma^{''}(s)
	\label{lambda2}
\end{equation}
where $\tau(s)$ is defined by:
\begin{equation}
	\tau(s) = \tilde{\tau}(s) + 2\pi(s)
	\label{tau}
\end{equation}
The solution to hypergeometric equation is given by Rodrigues relation:
\begin{equation}
	y_{n}(s) = \frac{B_{n}}{\tilde{\rho}(s)}\frac{d^n}{ds^n}[\sigma^{n}\tilde{\rho}(s)]
	\label{y}
\end{equation}
$B_n$ is the normalization coefficient and $\tilde{\rho}(s)$ must obey the following relation:

\begin{equation}
	(\sigma(s)\tilde{\rho}(s))^{'} = \tau(s)\tilde{\rho}(s)
\end{equation}

Then

\begin{equation}
	\frac{\tilde{\rho}^{'}}{\tilde{\rho}} = \frac{\tau - \sigma^{'} }{\sigma}
\end{equation}

 We have the following solution:

\begin{equation}
	\tilde{\rho}(s) = \exp\left(\int \frac{\tau - \sigma^{'} }{\sigma}\;ds\right)
	\label{sol}
\end{equation}
\textcolor{black}{
\section{\label{estim}Estimate of $\boldsymbol{\lambda_{\theta,\phi, \theta \phi}}$ \label{blast}}}
The blast wave model, described in detail in \cite{blast1}, is to parametrize the transverse plane flow profile of the fireball to obtain soft observables and relate them to particle properties via a Cooper-Frye type formula \cite{zanna}.
The fireball is assumed to be an ellipse in coordinate space
\begin{equation}
	x = r_{max}\sqrt{1-\epsilon} \cos\phi,\;\;\;\;\;\;\; y = r_{max}\sqrt{1+\epsilon} \sin\phi,
\end{equation}
where the $r_{max}$ and $\epsilon$ represents the parameters of model and $\phi$ is the azimutal angle in the transverse plane. To obtain the thermal vorticity, we need first define the fluid velocity that describes the azimuthally asymmetric fluid flow, parametrized by a Hubble-type Ansatz
\begin{equation}
	u^{\mu} = \left(\frac{t}{N},\frac{x}{N}\sqrt{1+\tilde{\delta}},\frac{y}{N}\sqrt{1-\tilde{\delta}}, \frac{z}{N}\right)
\end{equation}
Where the normalization factor is equal to $N = \sqrt{\tau^2-(x^2-y^2)\tilde{\delta}}$ and the proper time is $\tau^2 = t^2 - x^2 - y^2 - z^2$ ensuring, $u^{\mu}u_{\mu} = 1$.
At this point, we can obtain the thermal vorticity from the equation eq6 defined above:
\begin{equation}
	\omega_{\mu\nu} = -\frac{1}{2T}(\partial_{\mu} u_{\nu} - \partial_{\nu} u_{\mu}) - \frac{1}{2T^2}(u_{\mu}\partial_{\nu} T  - u_{\nu}\partial_{\mu} T ).
\end{equation}
The components are given explicitly in \cite{blast1}. Now, we will suppose that the $\omega$ is defined in the following way:
\begin{equation}
	\omega = \sqrt{\omega^2_x + \omega^2_y + \omega^2_z}
	\label{blastOmega}
\end{equation}
Where the spacial components to vorticity are given by $\omega^i = \frac{1}{2} \epsilon^{ijk} \omega^{jk}$. Now, make the average on the azimuthal angle, $\phi$, using the parameters to centrality $30–60\%$ in midrapidity $\eta = 0$ in equation \ref{blastOmega}. We have:
\begin{equation}
	\left<\omega\right> = \frac{1}{2\pi}\int^{2\pi}_{0}\omega(r_{max},\epsilon,\tilde{\delta}) d\phi = 24.08\;MeV
	\label{blastOmegaNumeric}
\end{equation}
We can make an estimative for this value for thermal vorticity \ref{blastOmegaNumeric} using equation \ref{cir} in the following form:  $C = 2\pi\left<\omega\right> \; r_0^2$.
Now, using $r_0 = 1/\delta$, we have:
\begin{equation}
	C = \frac{2\pi\left<\omega\right>}{\delta^2}
	\label{constNum}
\end{equation}
Using the numerical parameters reported in \cite{blast1,blast2} via a fit to experimental data, 
we can obtain the vorticity.

To obtain an estimate of the maximum value of 
$\lambda_{\theta,\phi,\theta\phi}$ we also need a numerical estimate of $\theta_r,\phi_r$ in Eq. \ref{sys3}.  This can be obtained from inverting Fig. \ref{states}.
The final result, calculated by plugging in these values in equations \ref{charmcoeff}, \ref{constNum} and \ref{densityMatrix} is shown in Table \ref{tab_quarkonium1}.

This is a very rough estimate, since these parameters were fitted at a different energy, and since only the second Fourier coefficient of the flow is considered (in vorticity all coefficients mix non-trivially).   The latter reason means we cannot calculate the exact azimuthal dependence.  However, since $v_n$ changes slowly with energy we  are confident this is a good order-of-magnitude estimate of the effect experimentalists are looking for.    If azimuthal modulation is found, more realistic models will be needed to describe it quantitatively.
\section{Quarkonium wave function determination\label{solution}}
Now, we can obtain the wave function from the equation \ref{sol} and using the equation \ref{tau2} and $\sigma(x) = x^2$. From it, we get
\begin{equation}
	\tilde{\rho}(x) = C_1 x^{-H_1/\sqrt{H_0}}e^{-2\sqrt{H_0}/x}
	\label{rhoaux}
\end{equation}

We are able to obtain the value of function $\phi(x)$ from equation \ref{tauaux} and using the values of equation \ref{tauaux}. Then
\begin{equation}
	\frac{\phi^{'}(x)}{\phi(x)} = -\frac{1}{2\sqrt{H_0}}\left(\frac{H_1}{x}-\frac{2H_0}{x^2}\right)
\end{equation}
Then
\begin{equation}
	\phi(x) = C_2\;x^{-H_1/2\sqrt{H_0}}e^{-\sqrt{H_0}/x}
	\label{phiaux}
\end{equation}
We can determine the $y_n(r)$ from equations \ref{y}, \ref{rhoaux} and $\sigma(r)$. Then,
\begin{equation}
	y_n(r) = B_n x^{H_1/\sqrt{H_0}}e^{2\sqrt{H_0}/x}\frac{d^n}{dx^n}\left[x^{2n-H_1/\sqrt{H_0}}e^{-2\sqrt{H_0}/x}\right]
\label{y2}
\end{equation}
From equation \ref{phiaux}, \ref{y2} and putting in \ref{trans}, we have that
\begin{equation}
	\psi(x) = B_n x^{H_1/2\sqrt{H_0}}e^{\sqrt{H_0}/x}\frac{d^n}{dx^n}\left[x^{2n-H_1/\sqrt{H_0}}e^{-2\sqrt{H_0}/x}\right]
\label{psiaux}
\end{equation}
%We can rewrite the equation \ref{psiaux} using the Laguerre Polynomials \cite{NU,ARF}. Defining the Laguerre polinomial the following form:
%\begin{equation}
%	L^{\sqrt{1+4H_0}}_{n}(2\sqrt{H_2}\;r) = \frac{1}{n!}e^{-\sqrt{H_2}\;r} r^{-\sqrt{1+4H_0}}\frac{d^n}{dr^n}(e^{-2\sqrt{H_2}\;r}\;r^{\sqrt{1+4H_0}+n})
%\end{equation}

Thereby,changing the variable $x=1/r$ we can write the radial wave fuction from equation \ref{psiaux} and relation $\psi(r) = r R(r)$. So:

\begin{equation}
	R_n(r) = B_n\;r^{-(H_1/2\sqrt{H_0})}e^{\sqrt{H_0}r}\left[-r^2\frac{d}{dr}\right]^n \left[x^{2n-H_1/\sqrt{H_0}}e^{-2\sqrt{H_0}/x}\right]
	\label{rfunc}
\end{equation}
%From the normalization of radial equation \ref{r} $\int_{0}^{\infty}R(r)R^{*}(r)r dr^2 = 1$, we can obtatin the constant $B_n$
%\begin{equation}
%	B_n = \sqrt{\frac{(n+\sqrt{1+4H_0})!}{n!}(2n+\sqrt{1+4H_0}+1)} 
%\end{equation}
Thus, we can write the wave fuction that following form:
\begin{equation}
	\Psi_{n,l,m}(r,\theta,\phi) = R_n(r)Y^m_l(\theta,\phi)
    \label{r}
\end{equation}
Using the value of \ref{rfunc}, we get:
\begin{equation}
	\Psi_{n,l,m}(r,\theta,\phi) =B_n\;r^{-(H_1/2\sqrt{H_0})}e^{\sqrt{H_0}r}\left[-r^2\frac{d}{dr}\right]^n \left[x^{2n-H_1/\sqrt{H_0}}e^{-2\sqrt{H_0}/x}\right]Y^m_l(\theta,\phi)
\label{wave}
\end{equation}

\end{document}